\documentclass[epsfig, usenatbib]{mn2e}
\usepackage{epsfig}
\usepackage{natbib}
\usepackage{graphicx}
\usepackage{fixltx2e}
\usepackage{multirow}
\usepackage{amsmath}
\usepackage{amssymb}
\usepackage{url}
\title{The progenitors of present-day massive red galaxies up to $z\approx0.7$ - finding passive galaxies using SDSS-I/II and SDSS-III}

\author[Tojeiro et al.]{
 \parbox{\textwidth}{Rita Tojeiro$^1$\thanks{E-mail: rita.tojeiro@port.ac.uk},  
  Will~J.~Percival$^1$, 
  David~A.~Wake$^2$, 
  Claudia~Maraston$^1$, \\
  Ramin~A.~Skibba$^3$,
  Idit~Zehavi$^4$,
  Ashley~J.~Ross$^1$,
  Jon~Brinkmann$^{5}$,\\
  Charlie~Conroy$^6$,
  Hong~Guo$^4$,
  Marc~Manera$^1$, 
  Karen~L.~Masters$^{1,7}$,\\
  Janine~Pforr$^8$,
  Lado~Samushia$^1$, 
  Donald~P.~Schneider$^{9,10},$
  Daniel~Thomas$^1$,\\
  Benjamin~A.~Weaver$^{11}$,
  Dmitry~Bizyaev$^{10}$,
  Howard~Brewington$^{5}$,\\
  Elena~Malanushenko$^{5}$,
  Viktor~Malanushenko$^{5}$,
  Daniel~Oravetz$^{5}$,
  Kaike~Pan$^{5}$,\\
  Alaina~Shelden$^{5}$,
  Audrey~Simmons$^{5}$,
  Stephanie~Snedden$^{5}$
  } \vspace{3mm}\\
$^1$Institute of Cosmology and Gravitation, Dennis Sciama Building, University of Portsmouth,
Burnaby Road, Portsmouth, PO1 3FX, UK \\
$^2$Astronomy Department, Yale University, P.O. Box 208101, New Haven, CT 06520, USA \\
$^3$Steward Observatory, University of Arizona, 933 N. Cherry Ave., Tucson, AZ 85721, USA \\
$^4$Department of Astronomy and CERCA, Case Western Reserve University, 10900 Euclid Avenue, Cleveland, OH 44106, USA.\\
$^{5}$Apache Point Observatory, P.O. Box 59, Sunspot, NM 88349-0059, USA\\
$^6$Harvard-Smithsonian Center for Astrophysics, Cambridge, MA, USA\\
$^7$SEPnet, South East Physics Network (www.sepnet.ac.uk)\\
$^8$NOAO, 950 N. Cherry Ave, Tucson, AZ 85719, USA \\
$^9$Department of Astronomy and Astrophysics, The Pennsylvania State University, University Park, PA 16802 \\
$^{10}$Institute for Gravitation and the Cosmos, The Pennsylvania State University, University Park, PA 16802 \\
$^{11}$Center for Cosmology and Particle Physics, New York University, New York, NY 10003 USA\\
}

\def\gs{\mathrel{\raise1.16pt\hbox{$>$}\kern-7.0pt %
\lower3.06pt\hbox{{$\scriptstyle \sim$}}}}         %
\def\ls{\mathrel{\raise1.16pt\hbox{$<$}\kern-7.0pt %
\lower3.06pt\hbox{{$\scriptstyle \sim$}}}}         %

\newcommand{\vmax}{$V_{\rm max}$ }
\newcommand{\vmatch}{$V_{\rm match}$ }

\begin{document}

\maketitle

\begin{abstract}
{We present a comprehensive study of 250,000 galaxies targeted by the Baryon Oscillation Spectroscopic Survey (BOSS) up to $z\approx 0.7$ with the specific goal of identifying and characterising a population of galaxies that has evolved without significant merging. We compute a likelihood that each BOSS galaxy is a progenitor of the Luminous Red Galaxies (LRGs) sample, targeted by SDSS-I/II up $z\approx 0.5$, by using the fossil record of LRGs and their inferred star-formation histories, metallicity histories and dust content. We determine merger rates, luminosity growth rates and the evolution of the large-scale clustering between the two surveys, and we investigate the effect of using different stellar population synthesis models in our conclusions. We demonstrate that our sample is slowly evolving (of the order of $2\pm1.5\%$ Gyr$^{-1}$ by merging) by computing the change in weighted luminosity-per-galaxy between the two samples, and that this result is robust to our choice of stellar population models. Our conclusions refer to the bright and massive end of the galaxy population, with $M_{i 0.55} \lesssim -22$, and $M_* \gtrsim 10^{11.2}M_\odot$, corresponding roughly to $95\%$ and $40\%$ of the LRGs and BOSS galaxy populations, respectively.
Our analysis further shows that any possible excess of flux in BOSS galaxies, when compared to LRGs, from potentially unresolved targets at $z\approx 0.55$ must be less than $1\%$ in the $r^{0.55}-$band (approximately equivalent to the $g-$band in the rest-frame of galaxies at $z=0.55$). When weighting the BOSS galaxies based on the predicted properties of the LRGs, and restricting the analysis to the reddest BOSS galaxies, we find an evolution of the large-scale clustering that is consistent with dynamical passive evolution, assuming a standard cosmology. We conclude that our likelihoods give a weighted sample that is as clean and as close to passive evolution (in dynamical terms, i.e. no or negligible merging) as possible, and that is optimal for cosmological studies.  }

\end{abstract}


\title{The progenitors of LRGs to z$\approx$0.7}

\section{Introduction}  \label{sec:intro}

The Baryonic Oscillation Spectroscopic Survey (BOSS), part of the Sloan Digital Sky Survey III (SDSS-III), is an ambitious galaxy redshift survey which will determine the expansion rate of the Universe up to $z\approx0.7$ by measuring the baryonic acoustic oscillations (BAO) and redshift-space distortions (RSD) in the galaxy power spectrum (\citealt{EisensteinEtAl11}). At the end of the five-year observing program, BOSS will have mapped 1.5 million massive galaxies in 10,000 square degrees of sky, resulting in unprecedented volume and galaxy density. Forecasts indicate that BOSS will yield measurements of the redshift-distance relation $d_A(z)$ and of the Hubble parameter $H(z)$ to $1\%$ and $1.8\%$ at $z=0.35$ and $1\%$ and $1.7\%$ at $z=0.55$, respectively (at the 1-sigma confidence level, \citealt{EisensteinEtAl11}). Using one third of the data, \cite{ReidEtAl12} placed initial constraints on $H(z=0.57)$ and $d_A(Z=0.57)$ at the 2.8\% and 4.8\% level respectively, whereas \cite{Aadvark} constrained $D_V(z=0.57) \equiv [cz(1+z)^2d_A^2H^{-1}]^{1/3}$ to $1.7\%$. To achieve the survey's  ambitious goals systematic uncertainties in the data, modelling and methodology must be kept to a minimum, and be understood as best as possible (see \citealt{RossEtAl11b,RossEtAl12} for a study on data systematics in BOSS). 

A source of uncertainty in the modelling and measurement of the BAO is galaxy bias. Different populations of galaxies relate differently to the underlying matter density field, yielding different biases and often different scales that mark the regime over which a linear, deterministic and scale-invariant bias model is applicable. To a certain extent one can parametrise over this uncertainty, but nonetheless an interesting question remains concerning how much gain is possible if the bias modelling and evolution with redshift were well understood. 

The best candidate for a population of galaxies with a well understood bias evolution is a population that has been evolving with no or very little merging: the bias evolution is easily modelled using the \cite{Fry96} formalism (see also \citealt{TegmarkEtAl98}). Massive red galaxies are the prime candidates for such a population - they are composed mostly of old stellar populations (e.g. \citealt{MarastonEtAl09}), and their growth via merging since a redshift of two has been constrained to be small ($<10\%$) even if strictly non-zero (e.g. \citealt{WakeEtAl08}). Halo-modelling analyses of massive red galaxies have repeatedly revealed a highly biased population ($b\approx 2$) with a low satellite fraction ($5$ to $10\%$ of galaxies are satellites) - see e.g. \cite{ZehaviEtAl05a,WakeEtAl08, ZhengZEtAl09,WhiteEtAl11} - confirming their suitability for cosmological studies. Departure from a pure passive dynamical evolution history has been shown to have a dependence on luminosity and colour \citep{TojeiroEtAl10, TojeiroEtAl11}, and this opens up the possibility of weighting galaxies appropriately, so as to maximise the contribution of those that are more likely to have been passively evolving, and minimise the contribution of those that are less likely to have done so. The SDSS-I/II survey targeted LRGs using a mix of colour and luminosity selection cuts such as to follow the evolution of a passively evolving population of stars. BOSS targeting, however, is much less restrictive in terms of luminosity and colour (especially at $z>0.45$ - see Section \ref{sec:data}). It is therefore not true that one population is automatically composed of the evolved products from the other.

One of the goals of this paper is to identify, in BOSS, the most likely progenitors of lower redshift SDSS-I/II LRGs, and design a set of weights that allow a selection of the galaxies that are linked by the same evolutionary history. 

Our other major goal is to place quantitative constraints on the formation and recent evolution of present-day luminous red galaxies, which in broad terms constitute a subset (at large luminosities, or stellar masses) of what are typically called early-type galaxies (ETGs). 
Efforts towards understanding ETGs and their evolution can be split into two categories: those that focus on their stellar content, and those that primarily aim to constrain their dynamical evolution, or merging history. Studies have been performed based on (see also references within): the mass or luminosity function of central galaxies \citep{WakeEtAl06, BrownEtAl07, FaberEtAl07, CoolEtAl08}, and of their satellites \citep{TalEtAl12}; colour-magnitude diagram \citep{CoolEtAl06, BernardiEtAl10b}; photometry SED fitting \citep{KavirajEtAl09, MarastonEtAl09}; absorption line fitting to individual galaxies' spectra \citep{TragerEtAl00, ThomasEtAl05, ThomasEtAl10, CarsonEtAl10} or to stacked spectra \citep{EisensteinEtAl03, GravesEtAl09, ZhuEtAl10}; full spectral fitting \citep{JimenezEtAl07}; close-pair counts \citep{BellEtAl06,BundyEtAl09} and clustering \citep{ZehaviEtAl05a,ShethEtAl06, MasjediEtAl06, ConroyEtAl07a, WhiteEtAl07, BrownEtAl08, MasjediEtAl08, WakeEtAl08, TojeiroEtAl10, deProprisEtAl10}. There is general agreement in the overall picture: ETGs constitute a uniform population of galaxies; are dominated by old and metal rich stellar populations; their mean ages (either mass- or light-weighted) decrease with luminosity; and the most luminous occupy the more dense environments. There is, however, an increasing amount of evidence pointing towards some amount of recent star formation in intermediate-mass ETGs (see e.g. \citealt{SchawinskiEtAl07,KavirajEtAl07,SalimEtAl10}). This amount of star formation is not in conflict with the hierarchical model of structure formation, and \cite{KavirajEtAl10}, through evidence coming from small morphological disruptions in early-type galaxies, argue that it can be explained from the contributions from minor-mergers. 

On the clustering side, halo modelling is rapidly being established as a successful tool to learn about galaxy formation (see e.g. \citealt{ZehaviEtAl05b,ZhengZEtAl07,SkibbaEtAl09a,SkibbaEtAl09b,RossEtAl09,ZhengZEtAl09,RossEtAl10,TinkerEtAl10, WakeEtAl11}, and references within).  
It is a powerful approach that connects galaxies with the dark matter halos in which they reside, and which describes the distribution of a population of galaxies in terms of centrals and satellites, as well as their relative ratio, as a function of halo mass (which is well correlated with luminosity, see e.g. \citealt{SwansonEtAl08,CresswellEtAl09,RossEtAl11}). 
E.g., \cite{ZhengZEtAl07} use luminosity dependent galaxy clustering at different epochs and the expected growth of dark matter halos to infer a growth due to star formation between $z=1$ and the present day, after roughly taking into account growth due to the merging of centrals and satellites. This type of description of galaxy assembly can then be directly compared to predictions from semi-analytical simulations (see \citealt{ZehaviEtAl12}). 

More specifically, the dynamical passive model can be directly tested by a halo model type of analyses. By performing HOD modelling at two different redshifts, one can evolve the best-fit halo model fitted at one redshift to another, assuming passive evolution. Comparison of the best-fit halo models provides insight about the dynamical evolution of the sample, particularly in terms of satellite accretion and disruption. For most of the samples chosen, analyses show that a purely passive model would predict too many satellites at low redshift, and therefore some galaxies must merge or be disrupted (see e.g. \citealt{ConroyEtAl07a, WhiteEtAl07,ZhengZEtAl07, BrownEtAl08,WakeEtAl08, SeoEtAl08}).
Measurements of merger rates of massive galaxies vary significantly (see Table 4 in \citealt{TojeiroEtAl10} for a summary), but luminosity growth via merging seems confined to something between 3-20\% since $z\approx1$. 

It seems increasingly likely that the assembly history of massive galaxies is inexorably linked to the existence of intra-cluster light (ICL) - a diffuse and scattered stellar component that can account for 10-50\% of the stellar mass in clusters (see e.g. \citealt{FeldmeierEtAl04, MihosEtAl05,PurcellEtAl07,YangEtAl09}). A likely mechanism of its formation is the disruption of satellite galaxies when halos merge (see e.g. \citealt{ConroyEtAl07b, PurcellEtAl07, WhiteEtAl07,YangEtAl09} and discussions therein). A lack of conservation of light, or stellar mass, in galaxy mergers has implications for the interpretation of the evolution of the luminosity function and inferred merger histories. The fraction of light lost by a merging satellite to the intra-cluster medium remains largely unconstrained, with estimates at the large halo mass end from the studies cited above varying between 15\% and 80\%. In the present work we make no explicit allowances for the loss of light to the ICM when two galaxies merge, but we will argue that our results are robust to this effect, within the limitations of the models and data. 

In the work presented here we approach the problem of galaxy assembly from a new direction, opposite in ethos to that of \cite{ZhengZEtAl07}. We will use state of the art modelling of the stellar evolution of a sample of galaxies to directly quantify growth from star formation, and from that infer a galaxy-merger history. We compute a model for the stellar evolution of SDSS-II LRGs by decomposing their spectra into a series of star-formation and metallicity histories, as well as dust content. This allows us to make predictions of their colour and  magnitudes at any redshift. This information, when combined with the target selection information for BOSS galaxies, constrains the regions in colour and magnitude space in BOSS within which progenitors of LRGs are more likely to reside. We then compute a set of weights that depend on the predicted evolution of each galaxy across the two surveys, and up-weight the objects that are more likely to be in both samples. The analysis we present depends on underlying assumptions about stellar evolution, initial mass functions and dust modelling. We perform the full analysis using two different sets of assumptions, so as to give the reader an idea of the dependence our final results on this type of uncertainty. 

Isolating the likely progenitors of LRGs in BOSS is in itself no test of the merging history of the sample. Following on from our analyses in \cite{TojeiroEtAl10, TojeiroEtAl11b}, we test the evolution in the number and luminosity density of the galaxies between LRGs and BOSS, as a way to measure the amount of merging or luminosity growth between the two redshift surveys. We also use a luminosity-weighted two-point correlation function to further test the dynamical passive hypothesis - weighting the galaxies by luminosity produces a clustering statistic that, on large-scales, is less sensitive to galaxies within the sample merging.

This paper is organised as follows: in Section \ref{sec:data} we describe our two data sets, including targeting; in Section \ref{sec:stellar_ev}  we explain how we compute a stellar evolution model that describes the stellar evolution of all galaxies and spans a redshift range between 0.23 and 0.7; in Section \ref{sec:sample_matching} we use this stellar evolution model to compute a set of weights that allows us to construct optimal samples of galaxies at different redshifts and explore the evolution of LRGs in the BOSS volume; in Section \ref{sec:population_evolution} we compute merger-rates and average luminosity growth across the samples and, in Section \ref{sec:clustering}, we compute the large-scale clustering of each of our samples and compare to predictions from a purely passive model. Finally we discuss and summarise our conclusions in Section \ref{sec:discussion}. Where required we assume a flat $\Lambda$ cold dark matter (LCDM) cosmology with $\Omega_m = 0.266$, $\Omega_\Lambda = 0.734$ and $H_0 = 71$ kms$^{-1}$Mpc$^{-1}$.

\section{Data} \label{sec:data}

The Sloan Digital Sky Survey (SDSS) has imaged over one quarter of the sky using a
dedicated 2.5m telescope in Apache Point, New Mexico \citep{GunnEtAl06}. For details on
the hardware, software and data-reduction see \citet{YorkEtAl00} and
\citet{StoughtonEtAl02}. In summary, the survey was carried out on a
mosaic CCD camera \citep{GunnEtAl98} and an auxiliary 0.5m telescope for photometric
calibration. Photometry was taken in five bands: $u, g, r, i$ and $z$ \citep{FukugitaEtAl96}, and magnitudes corrected for Galactic extinction using the dust maps of \cite{SchlegelEtAl98}. BOSS, a part of the SDSS-III survey \citep{EisensteinEtAl11}, has mapped an additional $5,200$ square degrees of southern galactic sky, increasing the total imaging SDSS footprint to nearly $14,500$ square degrees, or just over one third of the celestial sphere. All of the imaging was re-processed and released as part of SDSS Data Release 8 \citep{AiharaEtAl11}.

In SDSS-I/II, Luminous Red Galaxies (LRGs) were selected for spectroscopic
follow-up according to the target algorithm described in
\citet{EisensteinEtAl01},  designed to follow a passive stellar population in colour and apparent magnitude space. In this paper we analyse the latest SDSS LRG
spectroscopic sample (Data Release 7, \citealt{AbazajianEtAl09}), which includes
around 180,000 objects with a spectroscopic footprint of nearly 8000
sq. degrees and a redshift range $0.15 <z < 0.5$. In SDSS-III,  the BOSS target selection extends the SDSS-I/II algorithm to target fainter
and bluer galaxies in order to achieve a galaxy number density of $3\times10^{-4}$ h$^3$ Mpc$^{-3}$ and increase the redshift range out to $z\approx0.7$. The spectroscopic footprint of the BOSS data used in this sample covers almost $3500$ sq. degrees of sky, and corresponds to the upcoming Data Release 9, which will mark the first spectroscopic data release of BOSS. 

The targeting algorithms make use of five different definitions of magnitudes as follows: 
\begin{itemize}
\item SDSS uber-calibrated model magnitudes \citep{PadmanabhanEtAl98}, computed using either an exponential or a DeVaucouleurs light profile fit to the $r$-band only, denoted here with the $_{mod}$ subscript; 
\item cmodel magnitudes, computed using the best-fit linear combination of an exponential with a DeVaucouleurs light profile fit to each photometric band independently, and denoted here with the subscript $_{cmod}$; 
\item point-spread function (PSF) magnitudes, denoted with a $_{psf}$ subscript, and computed by fitting a PSF model to the galaxy; 
\item petrosian magnitudes, computed from the petrosian flux (the flux measured within twice the Petrosian radius, in turn defined using the surface brightness of the galaxy, see \citealt{Petrosian76,StraussEtAl02}), and denoted here by a subscript $_p$; and finally 
\item fibre magnitudes, computed within a 2 arcsec aperture, and denoted by a $_{fib2}$ subscript.
\end{itemize}

In SDSS-I/II, redshift LRGs were selected using two different algorithms. Cut-I predominantly but not exclusively targeted lower redshift galaxies ($z\lesssim0.43$) using the following selection criteria:

\begin{eqnarray}
 r_{p} & <  &13.1 + c_\parallel \\
r_{p} &<& 19.2 \\
 c_\perp& <& 0.2 \\
 \mu_{r,p} &< &24.2 \text{ mag arcsec}^2 \\
r_{psf} - r_{model} & >& 0.3,
\end{eqnarray}
\noindent
where the two colours, $c_\parallel$ and $c_\perp$ are defined as

\begin{eqnarray}
c_\parallel &=& 0.7(g-r) + 1.2[(r-i) - 0.18] \\
 c_\perp &=& (r-i) - (g-r)/4 - 0.18.
\end{eqnarray}

Model magnitudes are used for the colour cuts, and petrosian magnitudes for the apparent magnitude and surface brightness constraints. Note that whereas petrosian magnitudes naturally fail to account for flux outside twice the petrosian radii, and whereas this fraction varies as a function of galaxy type (see e.g. \citealt{GrahamEtAl05}), here they are simply used to define a sample of galaxies. When computing luminosity densities we always use cmodel magnitudes. Cut II mostly but not exclusively targets LRGs at $z\gtrsim 0.4$ following:

\begin{eqnarray}
r_p &< &19.5 \\
c_\perp &>& 0.45 - (g-r)/6 \\
(g-r) &> &1.3 + 0.35(r-i) \\
\mu_{r,p} &<& 24.2 \text{ mag arcsec}^2 \\
r_{psf} - r_{model} &>& 0.5.
\end{eqnarray}

Two separate algorithms are necessary as the passive stellar population turns sharply in a $g-r$ vs $r-i$ colour plane, when the 4000\AA\ break moves through the filters. 

In SDSS-III, galaxies at $z \lesssim 0.43$ are predominantly but not exclusively targeted by
the LOZ selection algorithm, akin to Cut I above, but 
extended to fainter magnitudes. A LOZ galaxy must pass the following:
\begin{eqnarray}
  r_{cmod} &<& 13.6 + c_\parallel/0.3, \label{eq:sliding_cut}\\
  |c_\perp| &<& 0.2, \label{eq:track_cut}\\
  16 < r_{cmod} &<& 19.6,
\end{eqnarray}
where the two auxiliary  colours $c_\parallel$ and $c_\perp$ are defined as for SDSS-I/II above. 

Galaxies at $z \gtrsim 0.43$ are predominantly but not exclusively targeted by
the CMASS selection algorithm, which extends the Cut II above by targeting both fainter and bluer galaxies. A CMASS
galaxy must pass the following criteria:
\begin{eqnarray}
  17.5 < i_{cmod} &<& 19.9, \\
  r_{mod} - i_{mod} &<& 2, \\
  d_\perp &>& 0.55, \\
  i_{fib2} &<& 21.5, \\
  i_{cmod} &<& 19.86 + 1.6(d_{perp} - 0.8),
\end{eqnarray}
where the auxiliary colour $d_\perp$ is defined as
\begin{equation}
  d_\perp = r_{mod} - i_{mod} - (g_{mod} - r_{mod})/8.0.
\end{equation}

CMASS objects must also pass the following star-galaxy separation cuts:
\begin{eqnarray}
  i_{psf} - i_{mod} &>& 0.2 + 0.2(20.0 - i_{mod}),\\
  z_{psf} - z_{mod} &>& 9.125 - 0.46z_{mod},
\end{eqnarray}
unless they also pass the LOZ cuts. 

\begin{figure}
\begin{center}
\includegraphics[width=3.5in]{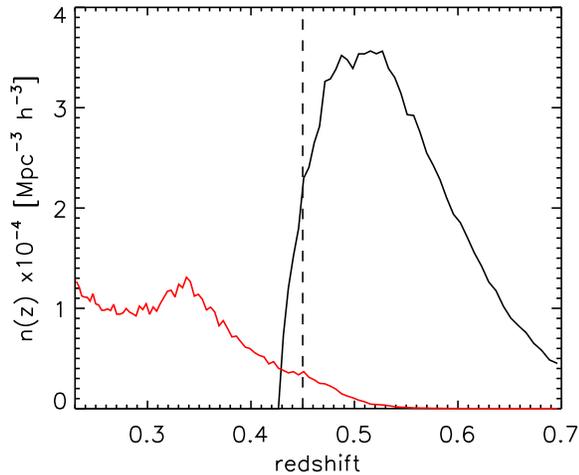}
\caption{Number density as a function of redshift for the LRG (red) and the CMASS (black) samples. The dashed line at $z=0.45$ shows our chosen hard boundary between the two surveys - we do not use any LRGs with $z>0.45$ nor any CMASS galaxies with $z<0.45$}
\label{fig:n_z}
\end{center}
\end{figure}

The CMASS selection algorithm was designed to loosely follow a constant stellar mass limit and,
unlike Cut-II in SDSS-II, it does not exclusively target red
objects. Therefore, whereas both the LRG and CMASS samples are
colour-selected, CMASS is a significantly more complete sample than the LRGs, especially at the bright end. In this paper we will split our data into two distinct redshift slices, with our lower redshifts slice ranging between $0.23 < z < 0.45$ and our higher redshift slice between $0.45 < z < 0.7$. Our low redshift slice consists exclusively of SDSS-I/II LRGs (Cut-I and Cut-II) and contains approximately 89,000 galaxies, and our high redshift slice consists exclusively of SDSS-III CMASS galaxies, with over 250,000 objects. The low redshift cut-off is motivated by our previous analysis of the LRGs that indicates the sample is significantly contaminated at lower redshifts (\citealt{TojeiroEtAl11, TojeiroEtAl11b}). We do not make use of LOZ galaxies for the main analysis presented in this paper, mainly due to the fact that the volume and number density sampled by LOZ currently lags behind that of the CMASS due to problems in target selection at the beginning of the observing run. We use LOZ galaxies only in Section \ref{sec:unresolved_pairs}, when investigating potentially unresolved targets in CMASS. The $n(z)$ distribution of our two samples is shown in Fig.~\ref{fig:n_z}.

\section{The stellar population modelling}\label{sec:stellar_ev}

We use the 124 stellar evolution models computed in \cite{TojeiroEtAl11} by stacking LRG spectra according to their luminosity, redshift and colour, and subsequently analysed them with VESPA (\citealt{TojeiroEtAl07,TojeiroEtAl09}) to obtain detailed star-formation histories as a function of lookback time. VESPA fits a linear combination of stellar populations of different ages and metallicities, modulated by a dust extinction, to the stacked optical spectra.  Each star-formation history can then be translated into a detailed evolution of any magnitude and colour with cosmic time. We have made no changes to these publicly available models other than increasing the sampling in redshift, to provide better resolved colour and magnitude evolution\footnote{The models from Tojeiro et al. (2011) are available at  \url{http://www.icg.port.ac.uk/~tojeiror/lrg_evolution/}}. We consider the solutions obtained with two sets of stellar population models: the Flexible Stellar Population Synthesis (FSPS) models of \cite{ConroyEtAl09} and \cite{ConroyAndGunn10}, and the stellar population models of \cite{MarastonEtAl11} (M11) - we refer the reader to Section 4 of \cite{TojeiroEtAl11} for detailed information on the differences and similarities between the two sets of assumptions, and we note that the most significant difference arises from the stellar evolution tracks\footnote{Briefly, notable differences lie in the choice of the shape of the initial mass function (IMF), isochrone tracks and stellar libraries. In the case of M11, we use a combination of a \cite{Kroupa01} IMF, the MILES stellar library of \cite{SanchezBlazquezEtAl06} and isochrones from \cite{CassisiEtAl97} and \cite{SchallerEtAl92} combined with the fuel consumption approach of \cite{RenziniAndBuzzoni86} for post Main-Sequence phases. For FSPS models we use a combination of a \cite{Chabrier03} IMF, with a MILES stellar library and the Padova isochrones of \cite{MarigoEtAl07,MarigoEtAl08}}. One of the main results in \cite{TojeiroEtAl11} is that, even though FSPS and M11 provide star-formation histories that have very similar mass-weighted ages that decrease with luminosity, in the M11 case this is due to the presence of a population of stars of young to intermediate ages (1-3 Gyr), whilst in the FSPS case this is due to a slightly younger main burst of star formation, which extends to lower ages with decreasing luminosity. These differences in the star-formation histories will have an impact on the results, and we will compare results obtained using both models throughout the paper. In Section \ref{sec:physical_model} we describe the star-formation histories recovered with both sets of models in detail.

\begin{figure*}
\begin{center}
\includegraphics[width=4.5in, angle=90]{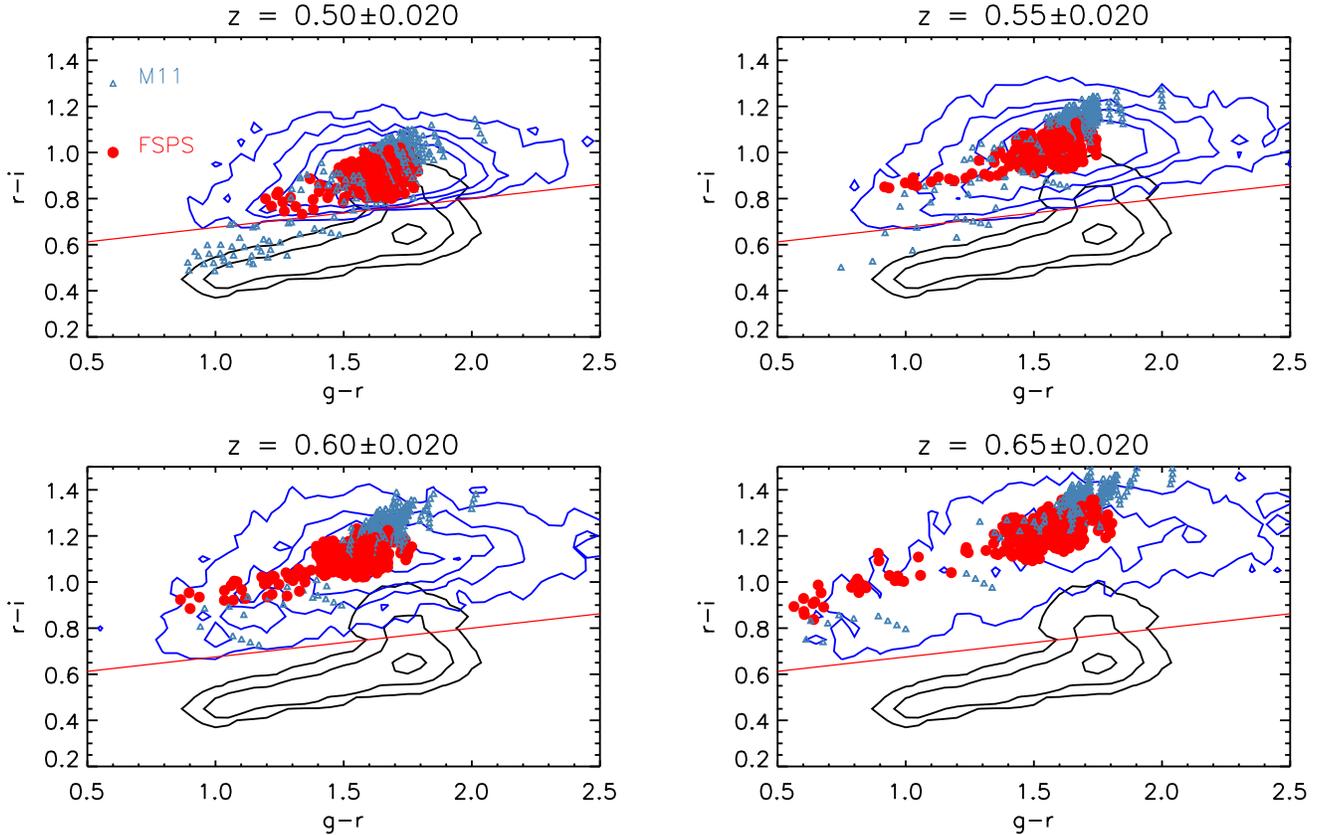}
\caption{The observed colour evolution of CMASS galaxies contrasted with the predicted colour evolution of LRGs at CMASS redshifts. In each panel the black contours show the number density of the full LRG sample in the $g-r$ vs $r-i$ plane. The blue contours show the number density of CMASS galaxies for a given redshift range (given for each panel). The red dots show the predicted colours of the LRGs at the same redshifts given by the FSPS models, and the blue dots show the predicted colours using the M11 models. The different dots correspond to the prediction of LRGs of different luminosity, colour and redshift. The solid red line shows the $d_\perp=0.55$ cut for reference.}
\label{fig:colour_ev}
\end{center}
\end{figure*}

In Fig.~\ref{fig:colour_ev} we show the $g-r$ and $r-i$ colours predicted by the fits to LRGs based on the different models (red dots for FSPS and blue for M11) and how they compare to the colours of observed CMASS galaxies in four redshift ranges (blue contours). The locus of the models traces the locus of the observed galaxies remarkably well. Furthermore, the FSPS models predict a tendency to have bluer colours with increased redshift, and that is tentatively matched by the data. M11 models follow broadly the same trend, with the main differences seen at $z=0.55$, where M11 models predict significantly bluer galaxies (some models predict a crossing of the $d_\perp$ cut). 

\subsection{The composite model}\label{sec:composite_model}
\begin{figure}
\begin{center}
\includegraphics[width=3.7in]{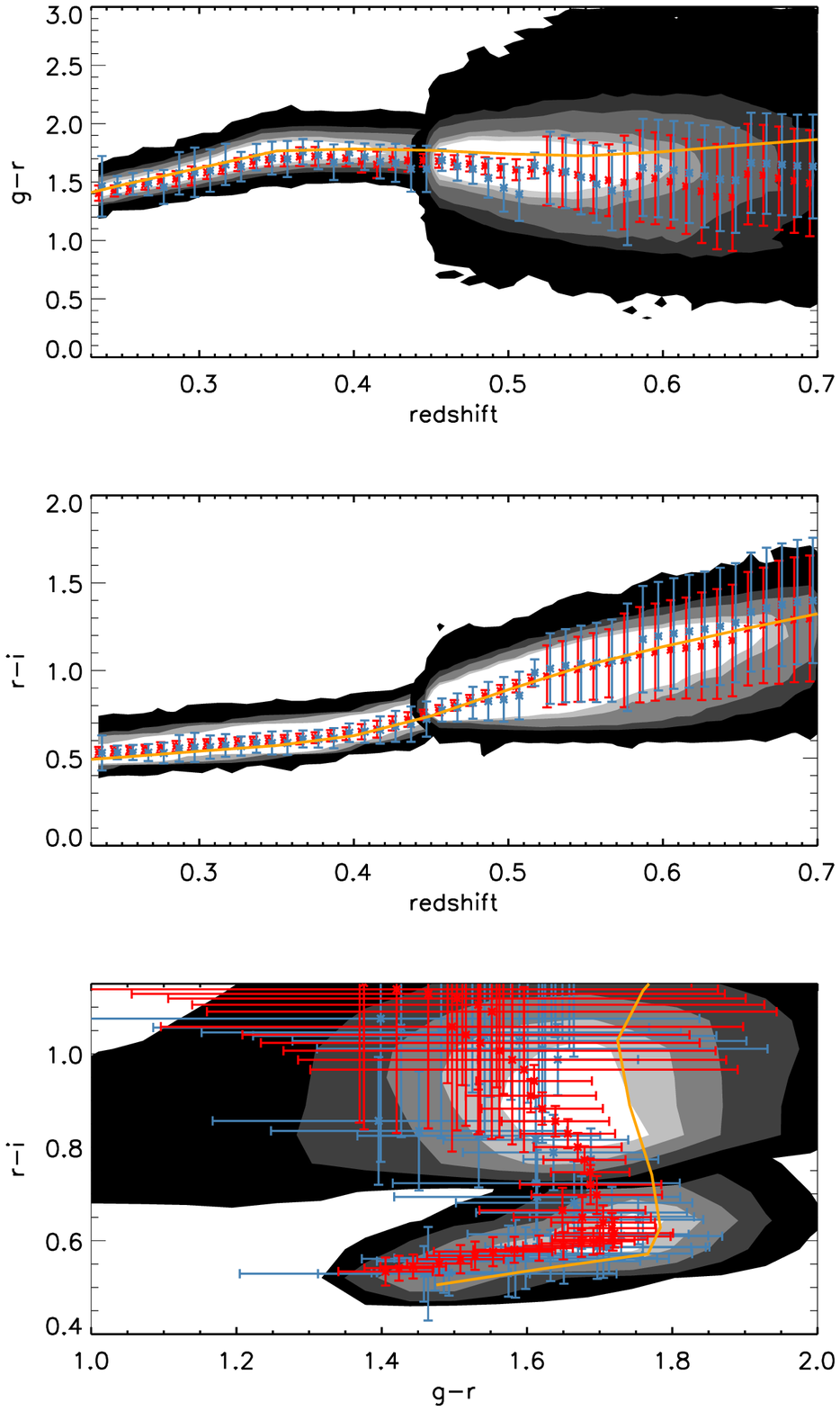}
\caption{The composite stellar evolution model, computed according to the procedure in Section~\ref{sec:composite_model}. In all panels the shaded contours show the number density of LRGs (at $z<0.45$ and on the bottom half of the last plot) and CMASS galaxies (at $z>0.45$ and on the top half of the last plot). The red (blue) solid line shows our composite stellar model obtained using the FSPS (M11) VESPA star-formation histories. It is a weighted average of the models shown in Fig.~\ref{fig:colour_ev}. The error bars show the 1$\sigma$ dispersion of the models shown in Fig.~\ref{fig:colour_ev} in each redshift bin. For reference, the yellow line shows the LRG purely passive model of \protect\cite{MarastonEtAl09} - see Section \ref{sec:passive_model}.}
\label{fig:composite_model}
\end{center}
\end{figure}
The clear advantage of our set of models is that it gives a data-driven grasp on the stochasticity of the population properties. We do not need to assume all targeted LRGs are the same and natural scatter in the colours - given by changes in metallicity and star-formation rate - can be trivially accounted for. We are limited in the sense that we can only predict the evolution of any galaxy to redshifts greater than the one it is observed at; this is because the fossil record can only hold information on the {\it past} history of a galaxy. Evolving a galaxy forward requires assumptions about any subsequent star-formation, or lack of it. In order to match the samples we need a stellar evolution model that spans the redshift range of both samples combined, and that we can use to evolve any galaxy to any redshift with minimal assumptions about their stellar evolution. We choose an approach where we compute a {\it single} weighted composite model that spans the redshift range of the sample, based on the 124 individual stacks. At each redshift we compute a mean spectrum, weighted by the number of galaxies that make a prediction for that particular redshift (i.e. observed at $z\le z'$). From this mean spectrum we compute a new set of magnitude and colours, that define what we will call our {\it composite model}. The k+E corrections of the composite model are the weighted means of the individual k+E corrections - this composite model is therefore our best estimate of the overall average colour and magnitude evolution of the full LRG sample. Note that this approach is formally the equivalent to taking the weighted mean of the 124 star formation histories for each stack, and using that weighted star formation history to recover the composite spectrum and correspondent models.

\begin{figure}
\begin{center}
\includegraphics[width=3.5in]{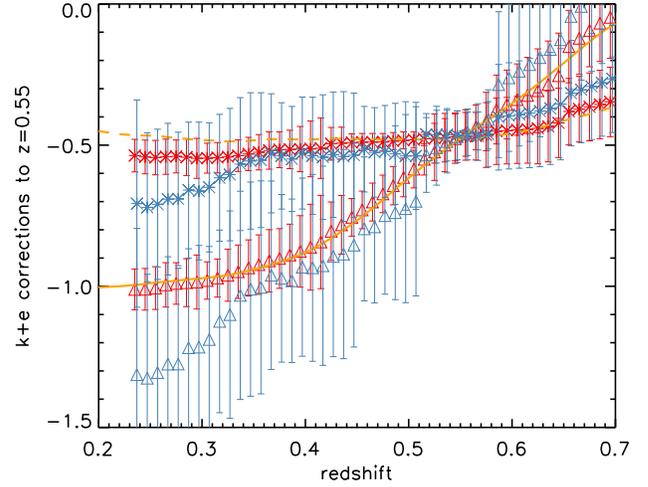}
\caption{K+e corrections in the $r_{0.55}$-band (triangles) and in the $i_{0.55}$-band (asterisks).The red lines refer to the FSPS models, and the blue lines to the M11 models. The error bars show the $1\sigma$ scatter around the mean from the 124 individual stacks. These corrections allow us to compute the evolved absolute magnitude of any galaxy at $z=0.55$, in the two shifted filters (therefore for galaxies at $z=0.55$ this correction is fixed and independent of their spectra or modelling). The corrections in the $r_{0.55}$ band are steeper because it traces the 4000\AA\ break at these redshifts - see Fig.~\ref{fig:filters}. The scatter in the M11 k+E corrections is larger, as these models predict stochastic events of star-formation at young to intermediate ages in some of the stacks. For reference, the yellow line shows the LRG purely passive model of \protect\cite{MarastonEtAl09}, as a dashed line for the $i_{0.55}$-band and as a solid line for the $r_{0.55}$-band - see Section \ref{sec:passive_model}.}
\label{fig:ke_corrections}
\end{center}
\end{figure}

We show the colour evolution of our model in Fig.~\ref{fig:composite_model}, and the K+e corrections in Fig.~\ref{fig:ke_corrections} (red for FSPS models, and blue for M11). These models are used to describe all galaxies in the study: LRGs and CMASS galaxies alike. For completeness in Section \ref{sec:beyond_composite} we briefly discuss the impact of using the strictly passive stellar evolution model of \cite{MarastonEtAl09}, or the full range of 124 individual models, on our results.

\subsubsection{Physical model}\label{sec:physical_model}

As mentioned in the previous section, VESPA solutions with the two different stellar population models give physical models for the galaxies that are qualitatively different, especially for LRGs at $z<0.25$. FSPS produces a model that is nearly completely passive, with less than $2-3\%$ by mass in stars that are younger than 3 Gyrs. M11 gives a model that sees over $90\%$ of the stellar mass formed over 12 Gyrs ago (for a galaxy at $z=0$), but which often puts a non-negligible amount of stars at ages of 1-3 Gyrs (up to $10\%$ in mass). This generates more scatter in the blue points in Fig.~\ref{fig:colour_ev} and, as a direct consequence, a larger scatter in Figs.~\ref{fig:composite_model} and \ref{fig:ke_corrections}. 

Small but non-negligible amounts of star-formation act to steepen the luminosity evolution (given by the k+E corrections), as the galaxy effectively 'loses' stars as we step back in redshift. More generally, a change in k+E corrections can also arise from different assumptions in the stellar evolution models, or from a different slope - or an evolving slope - of the initial mass function (IMF). In \cite{TojeiroEtAl10} we investigated the effects of an added redshift-dependent term to the k+E corrections, being motivated at the time by uncertainties in the slope of the IMF. Here we will perform no such investigation, but having two models with two different slopes for the k+E corrections provides an estimate of the impact of this uncertainty on our final results.

Both stellar population models give a constant metallicity with redshift, although M11 solutions are slightly more metal rich at $Z \approx 0.03$, whereas FSPS prefers a solution with $Z \approx 0.025$. 

Finally, the dust content is very similar in both cases - extinction increases with decreasing luminosity, increasing redshift and increasing $r- i$ colour, varying between $\tau_V = 0.2$ and $\tau_V=0.8$. Here $\tau_V$ is the optical depth at $\lambda = 5500\AA$ and the dust extinction is modelled according to a \cite{CharlotFall00} mixed-slab geometry (see \citealt{TojeiroEtAl11} for full details). The weighted average, and the effective extinction for the composite model, is  $\tau_V \sim 0.4-0.5$.

\subsection{K+e corrections}\label{sec:ke_corrections}
We follow closely the procedure of \citet{TojeiroEtAl10}, which we summarise here for completeness.

\begin{figure}
\begin{center}
\includegraphics[width=3.5in]{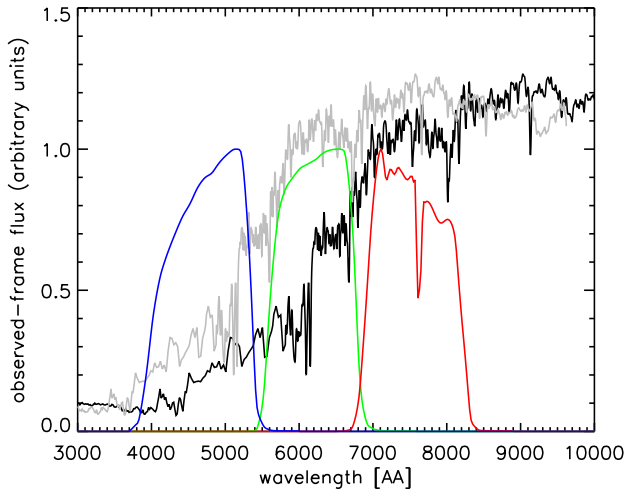}
\caption{The expected observed spectrum of a typical galaxy in the sample at $z=0.55$ (black). The three broadband filters used for target selection are overplotted: $g-$band in blue, $r-$band in green and $i-$band in red. For reference, we show in grey the expected observed spectrum of a galaxy at $z=0.3$.}
\label{fig:filters}
\end{center}
\end{figure}

Our composite model provides
$L_\lambda(t_{age})$, the luminosity per unit wavelength of a galaxy of
age $t_{age}$. We
K+e-correct all galaxies to a common redshift of $z_c=0.55$, and
calculate corrected absolute magnitudes in filters shifted to $z_c=0.55$ as
\begin{equation}\label{eq:abs_mag}
  M_{r 0.55} = r_{cmod} - 5\log_{10} \left\{ \frac{D_L(z_i)}{10 \mathrm{pc}} \right\}
  - Ke(z, z_c),
\end{equation}
with
\begin{eqnarray} \label{eq:ke_corrs}
  \lefteqn{Ke(z, z_c) =}    \\
  &= -2.5 \log_{10} \left\{ \frac{1}{1+z} \frac{ \int T_{\lambda_o}
      L_{\lambda_o} (z) \lambda_o d\lambda_o \int T_{\lambda/(1+z_c)}
      \lambda_e^{-1} d\lambda_e} {\int T_{\lambda_o/(1+z_c)}
      L_{\lambda_e}(z_c) \lambda_e d\lambda_e \int
      T_{\lambda_o}\lambda_o^{-1} d\lambda_o} \right\}. \nonumber
\end{eqnarray}
$\lambda_o$ is in the observed frame and $\lambda_e$ in the emitted
frame. $T_\lambda$ is the SDSS's $r$-band filter response, and
$L_\lambda(z)$ the luminosity density of a galaxy at redshift $z$, given the
fiducial model. We also compute $M_{i 0.55}$, using exactly the same procedure on the $i$-band. Note that for a galaxy at $z=0.55$, the K+e correction is independent of the observed or modelled spectrum and equals $-2.5 \log_{10}\left(\frac{1}{1+z}\right)$. By choosing $z_c = 0.55$, roughly the peak of the redshift distribution of CMASS galaxies, we minimise the effect of the modelling on CMASS galaxies. The other option would have been to k+e correct to median redshift of LRGs. However, as the composite stellar population model is based on the spectra of LRGs, its predictions must be at least as robust for LRGs as for CMASS galaxies, if not more so. Therefore, our procedure is the more robust approach. We show the k+e correction in the $r-$ and $i-$bands in Fig.~\ref{fig:ke_corrections}. For reference, in Fig.~\ref{fig:filters} we show the expected observed-frame spectrum of a typical galaxy in the sample at $z_c=0.55$, along side the three broadband filters used in this paper.

\subsection{Comparing CMASS galaxies and LRGs}\label{sec:samples}

\begin{figure*}
\begin{center}
\includegraphics[width=4.5in, angle=90]{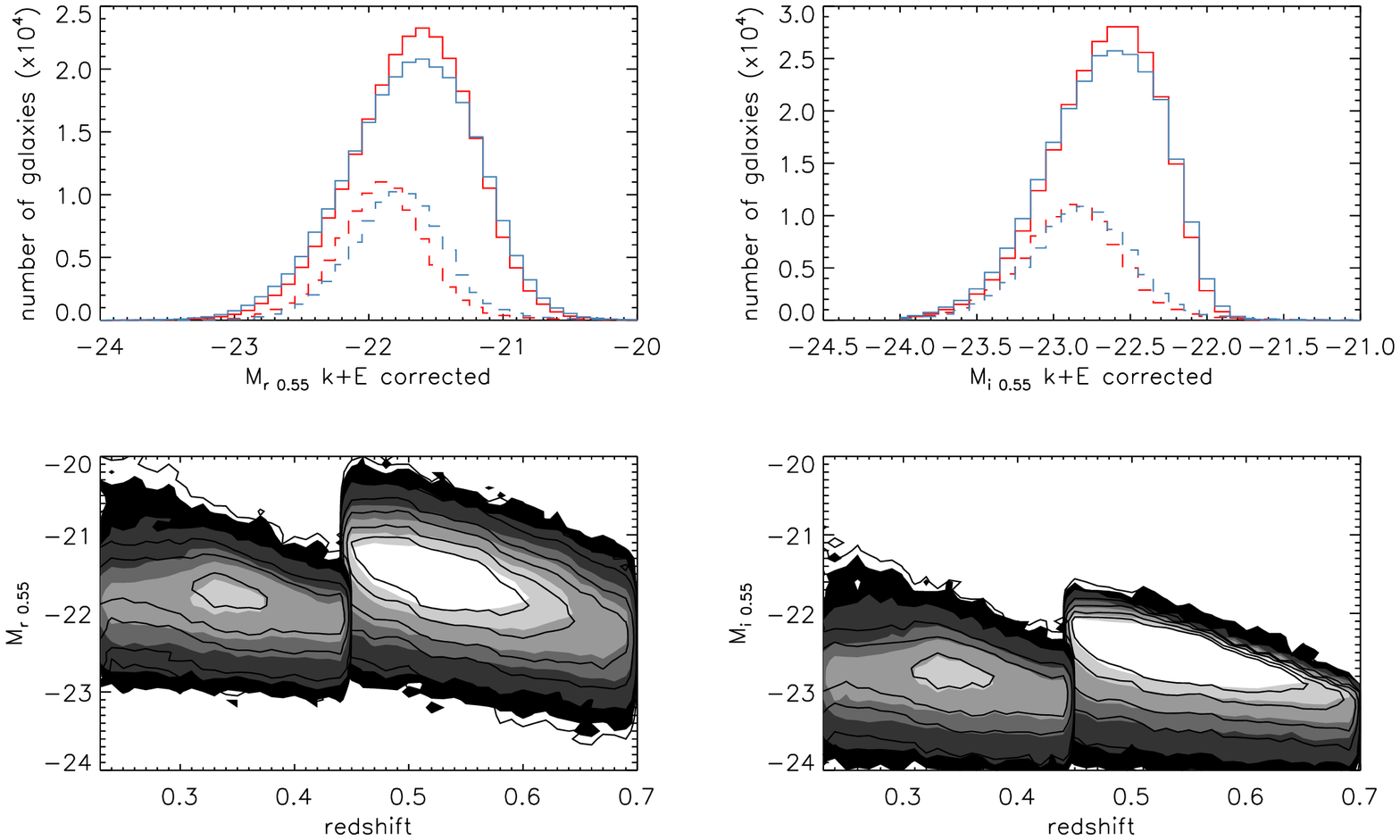}
\caption{Comparing K+e corrected magnitudes in SDSS-I/II LRGs and BOSS CMASS galaxies. {\it Top:} the distribution of absolute magnitudes for LRGs (dashed lines) and CMASS galaxies (solid lines). The different colours show the results from using different stellar population models, with FSPS in red and M11 in blue. The two panels show the magnitude computed either in the rest-frame $r-$ or $i-$band. {\it Bottom:} the absolute-magnitude with redshift on both samples. FSPS results are shown in the solid contours, and M11 in the line contours. The samples are split at $z=0.45$; we do not use any LRGs with $z>0.45$ nor any CMASS galaxies with $z<0.45$. These plots show clearly the  reach to fainter magnitudes of the CMASS sample. See main text for a discussion on the effect of the stellar population models.}
\label{fig:mag_comparisons}
\end{center}
\end{figure*}

We can use the K+e-corrected absolute magnitudes to broadly characterise the two samples. Fig.~\ref{fig:mag_comparisons} shows a simple comparison of the magnitude distributions for both samples and their evolution with redshift computed for both the $r-$ and the $i-$band. Once again we show the results for the FSPS model in red and for the M11 model in blue. Here the only model differences come through the K+e corrections, with the different slopes between models (shown in Fig.~\ref{fig:ke_corrections}) naturally giving different k+E corrected absolute magnitudes. M11 shows a steeper slope with respect to FSPS, with the crossing point at $z_c=0.55$. So for a galaxy at $z < 0.55$ M11 will predict a {\it fainter} k+E corrected magnitude at $z=0.55$. Conversely, for a galaxy at $z > 0.55$, M11 will predict a {\it brighter} k+E corrected magnitude at $z=0.55$. By construction, the magnitudes for galaxies sitting at $z=0.55$ will match for both models due to our choice of filters. The top panel of Fig.~\ref{fig:mag_comparisons} shows the effect of having different slopes for the k+E corrections - for LRGs this is about 0.3 magnitudes in the $r_{0.55}$-band; for CMASS galaxies it is much smaller, at less than 0.1 magnitudes. These values are roughly halved for the $i_{0.55}$-band.  The bottom two panels of Fig.~\ref{fig:mag_comparisons} show the evolution of the corrected magnitudes with redshift (solid contours for FSPS and line contours for M11). As expected, we see a steeper evolution with redshift using the M11 contours. 

\begin{figure*}
\begin{center}
\includegraphics[width=3.2in]{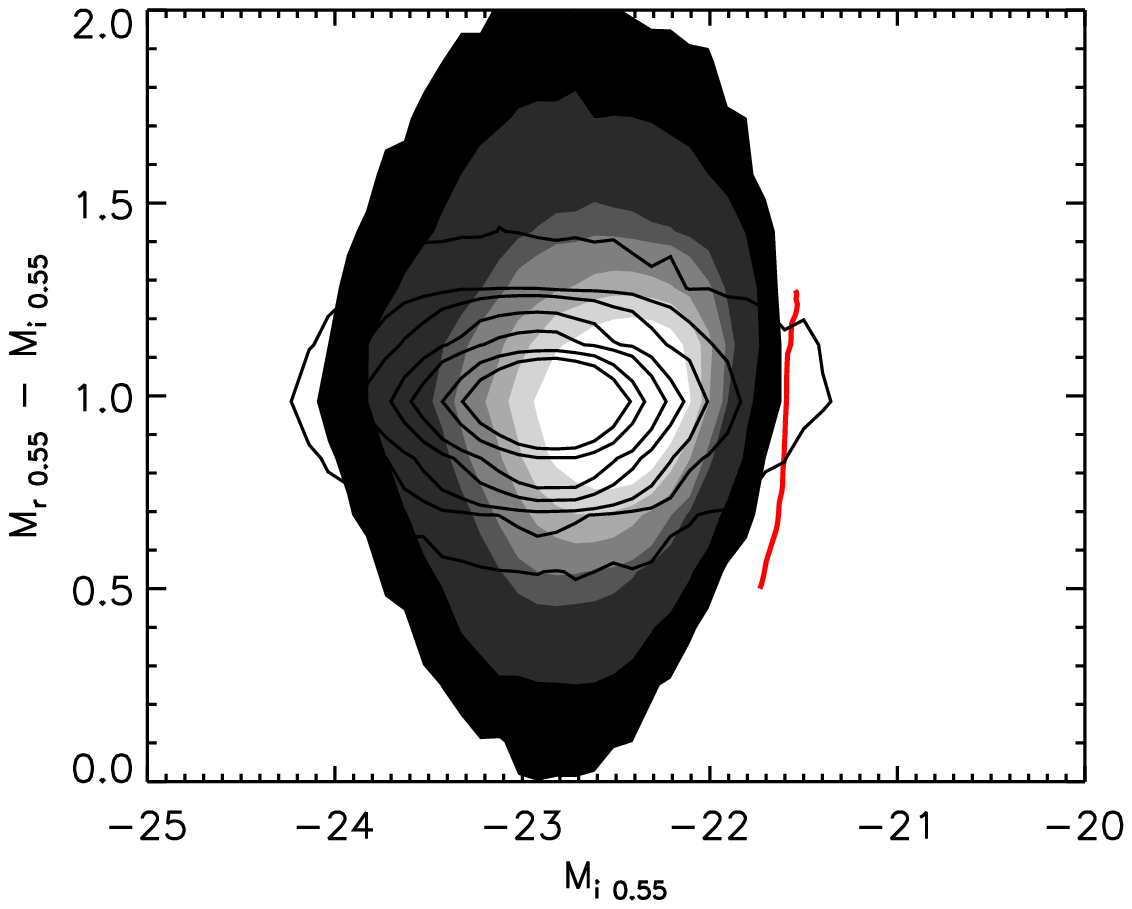}
\includegraphics[width=3.2in]{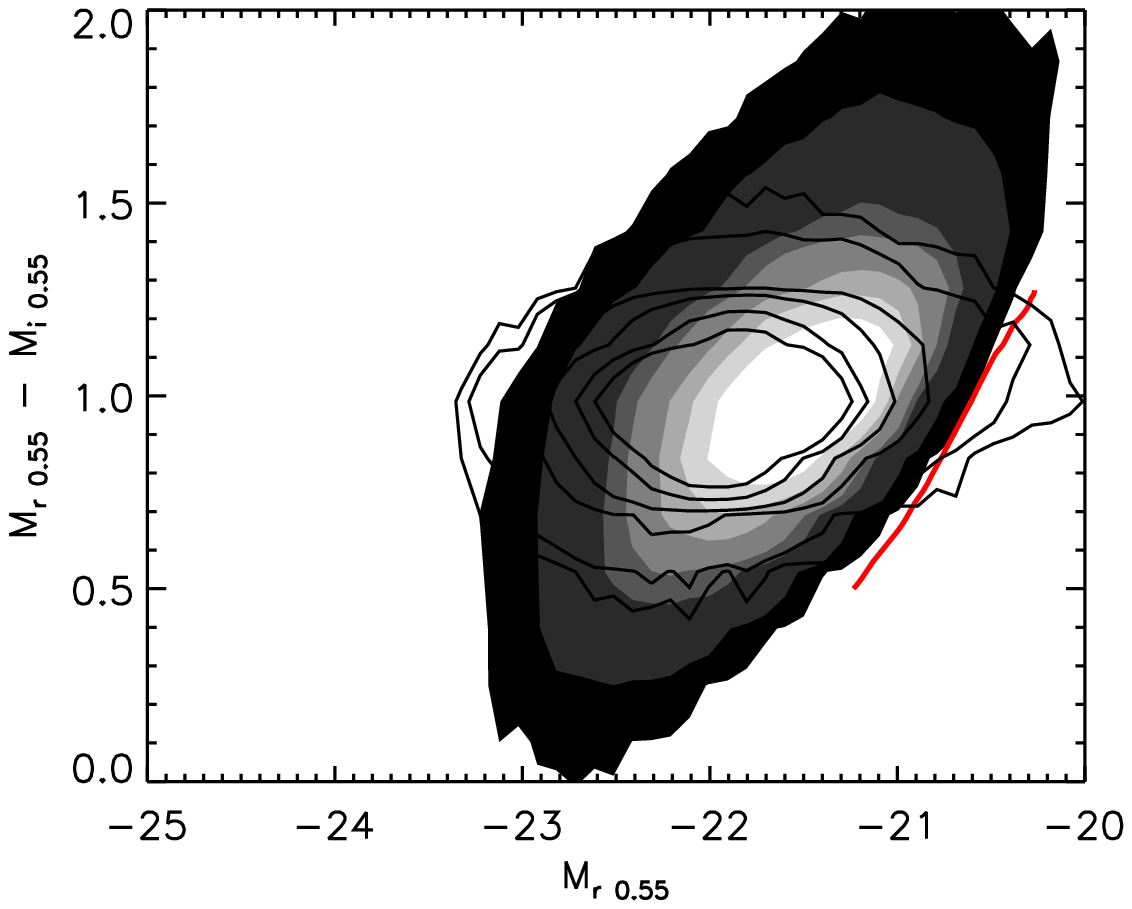}
\caption{Rest-frame, k+e corrected colour-magnitude relations for CMASS galaxies (filled contours) and LRGs (overploted black contours), as a function of $M_{i 0.55}$ shown on the left panel, and as a function of $M_{r 0.55}$ on the right hand side. CMASS galaxies show a broader range in their rest-frame $M_{r 0.55} - M_{i 0.55}$, as well as fainter reach and median in both magnitudes. The right-hand side plot shows a clear trend of rest-frame colour with $M_{r 0.55}$, with redder colour going with lower luminosity. This is trend is a result of target selection, particularly the magnitude cut - we show the expected evolution of the colour-magnitude relation of an object at the faint end of the survey (cmodel$=19.9$ at $z=0.45$) and an observed colour of $r-i=0.8$, between $z=0.25$ and $z=0.7$. Any object to the faint side of the red lines would fail the magnitude cuts of the CMASS algorithm.  }
\label{fig:colour_magnitude}
\end{center}
\end{figure*}

Fig.~\ref{fig:colour_magnitude} displays colour-magnitude relations. Here we show only the results using FSPS models as the results are similar in both cases. The CMASS sample has a broader range in absolute magnitude and colour than the LRG sample, as expected given the larger number density. The clear trend seen between rest-frame colour and $M_{r 0.55}$ is explained simply by target selection. To help make this point we show the expected evolution of the colour-magnitude relation of an object at the faint end of the survey (cmodel $=19.9$ at $z=0.45$) and an observed colour of $r-i=0.8$, between $z=0.25$ and $z=0.7$ - this is the red line in both plots. Any object to the faint side of the red lines would fail the magnitude cuts in the $i-$band of the CMASS algorithm. This gives an obvious artefact when plotting $M_{r 0.55}$ vs colour, where upon the CMASS selection does not select faint blue galaxies. The bright end slope is a consequence of volume effects, coupled with the slope of a typical galaxy spectra.



%

\section{Sample matching}\label{sec:sample_matching}
We now construct galaxy samples at high and low redshift that are coeval according to our composite stellar evolution models. We continue to closely follow the methodology of \cite{TojeiroEtAl10}, which we summarise below. 
We have to take into account three redshift-dependent effects:
\begin{enumerate}
  \item the intrinsic evolution of the colour and brightness of the galaxies;
  \item the varying errors on galaxy colour measurements; and
  \item the varying survey selection function.
\end{enumerate}

Our correction for (i) is given by our composite stellar evolution model. We include an evolving colour scatter term to allow for
(ii). \cite{TojeiroEtAl10} used the population scatter around the stellar evolution model with redshift. \cite{TojeiroEtAl11b} updated this term to be based on the evolution of photometric errors as a function of apparent magnitudes, which were modelled as a function of redshift - see their Section 3. The motivation was two fold: firstly the  photometric errors are driven principally by the apparent magnitude of an object, rather than its redshift; and secondly this is less dependent on choice of stellar evolution modelling. We adopt this approach here. For (iii) we construct a set of weights that assures a given population of
galaxies - in terms of colour and absolute magnitude - is given the
same weight in the high and low redshift samples, as described in the next section.

\subsection{Weighting scheme}\label{sec:weighting_scheme}

We use the weighting scheme of \cite{TojeiroEtAl10}, which keeps the total weight of each
{\it galaxy population} the same in different redshift slices. 

Suppose an LRG, $g_A$, is faint and therefore can only be seen in a small fraction of the CMASS volume, $f_V$, but can be seen in the full LRG volume. Then our weighting scheme will give $g_A$ a weight that is equal to $f_V$. Consider now a faint CMASS galaxy, that is observed in $f_V$, and whose magnitude and colour evolution matches those predicted for $g_A$. This galaxy will by definition also only be observed in a fraction $f_V$ of the CMASS volume. Our weighting scheme gives $g_B$ a weight on unity. Note this is the opposite approach to the traditional \vmax weight, which would {\it up-weight} $g_B$ by $1/f_V$ and give $g_A$ a weight of unity.

Explicitly,
for an LRG in a volume $V_{LRG}$ we calculate
\begin{equation}\label{eq:wa}
  V_{\rm{match},i} = \frac{V_{LRG}}{V^{LRG}_{\rm{max},i}} \times  \mathrm{min} 
    \left\{ \frac{V^{LRG}_{\rm{max},i}}{V_{LRG}}, \frac{V^{CMASS}_{\rm{max},i}}{V_{CMASS}} \right\},
\end{equation}
and similarly for a CMASS galaxy, in a volume $V_{CMASS}$:
\begin{equation}\label{eq:wb}
  V_{\rm{match},i} = \frac{V_{CMASS}}{V^{CMASS}_{\rm{max},i}}  \times \mathrm{min} 
    \left\{ \frac{V^{LRG}_{\rm{max},i}}{V_{LRG}}, \frac{V^{CMASS}_{\rm{max},i}}{V_{CMASS}} \right\}.
\end{equation}
where $V_{\rm{max},i}$ is the volume a galaxy $i$ would have been observed in either survey, according to the full target selection cuts and the evolution of its colour and magnitude, as given by the composite model.

Where the traditional $V_{\rm max}$ estimator would up-weight
galaxies only visible in a fraction of the volume they were observed in, we instead give these galaxies a weight of unity and down-weight the corresponding galaxies with the same properties observed in the other volume.

The interpretation of the \vmatch weight is different than that of the traditional \vmax
weighting. Whereas the latter gives us the means to correct for
incompleteness and yields true space densities, the former must be interpreted as a weighting scheme rather than a completeness
correction. I.e., \vmatch weighted number and luminosity densities {\em are still potentially volume incomplete},
but the populations are weighted in such a way that they are equally
represented at both redshifts. We can compare the distribution of
total weighted luminosity for the two slices, but we cannot interpret
these functions as giving the true luminosity density. 

The advantage of this weighting scheme is that we sample different populations equally based on volume, and therefore obtain a weighted population such that galaxies observed throughout a large volume are up-weighted. It also implicitly checks that we are only using populations that exist in both samples, without having to do such a test explicitly (e.g. \citealt{WakeEtAl06}).


\subsection{The progenitors of LRGs}\label{sec:progenitors}

\begin{figure*}
\begin{center}
\includegraphics[width=3.2in]{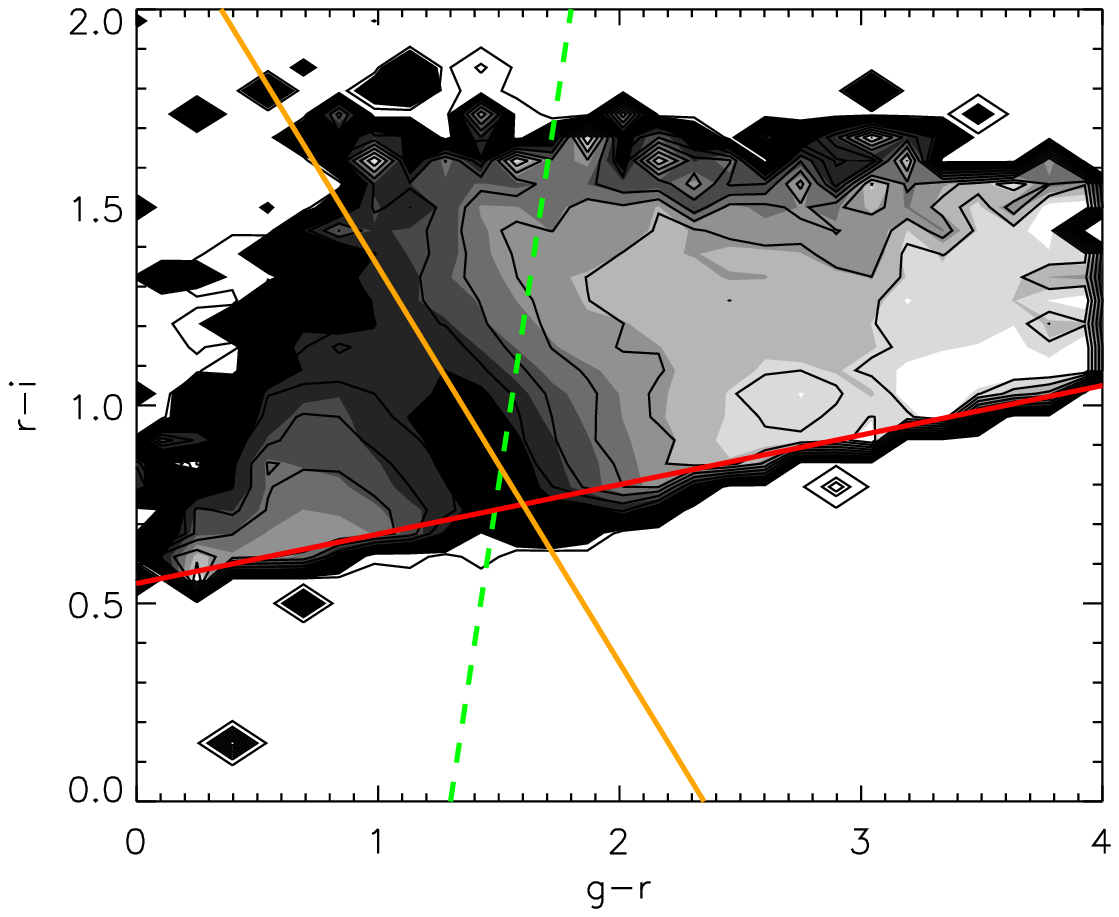}
\includegraphics[width=3.2in]{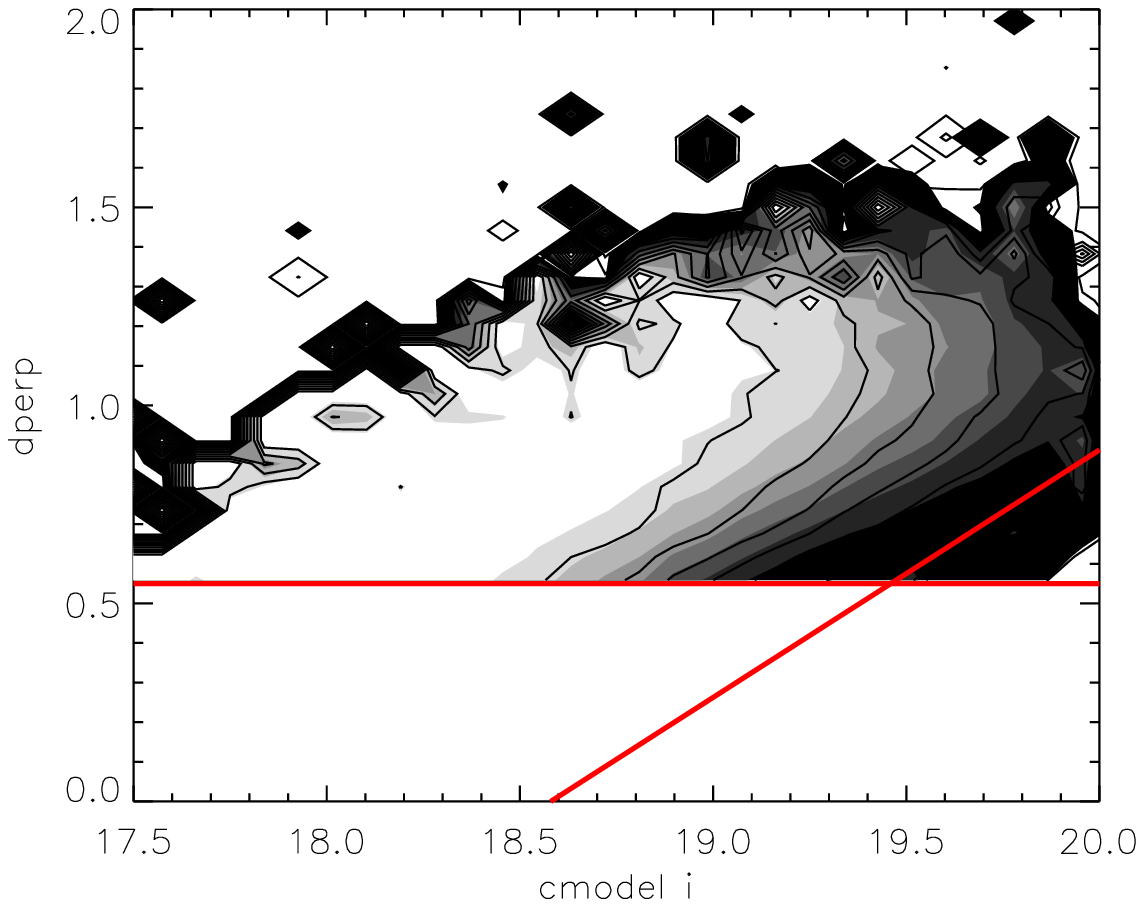}
\caption{Average \vmatch weight as a function of colours and $i-$band magnitude, shown for the two main targeting parameter space diagrams in CMASS. A darker colour corresponds to a lower value of \vmatch, and the brighter colours to the regions in parameter space that have the largest likelihood of being progenitors of the LRG sample. The red solid lines show targeting cuts. The orange line on the plot on the left shows the morphology cut derived in \protect\cite{MastersEtAl11}, and the dashed green line shows the blue cut of the cut-II selection in \protect\cite{EisensteinEtAl01}. }
\label{fig:Vmatch_density}
\end{center}
\end{figure*}

A large value of \vmatch (\vmatch varies between 0 and 1) indicates that a galaxy belongs to a population that can be observed across a large fraction of both surveys, and a small value of \vmatch means a population of galaxies is only present in a small fraction of the volume in at least one of the surveys. In other words, the larger this value for a CMASS galaxy, the more likely this galaxy is a progenitor of a typical LRG galaxy, and vice-versa. 

Fig. \ref{fig:Vmatch_density} shows a mapping of the average value of this weight onto the two CMASS targeting parameter spaces: a $g-r$ vs $r-i$ plot, and a $d_\perp$ vs the cmodel magnitude in the $i$-band. We show the results using the FSPS models in the solid contours and the results using M11 in the line contours, which are qualitatively  similar. The colour-colour plot shows a clear trend for the average value of \vmatch to increase to redder $g-r$ colours, as expected if LRGs were exclusively made of metal rich and old stars. Interestingly, we also see that some blue regions of the colour-colour plot display an increase of the average value of the \vmatch weight. This relation is a result of the small but significant amounts of young to intermediate-aged stars detected in LRG spectra at BOSS redshifts (corresponding roughly to stars aged between 1 and 3 Gyr in SDSS-I/II galaxies). The orange line in the left-hand plot of Fig.~\ref{fig:Vmatch_density} shows the $g-i = 2.35$ cut of \cite{MastersEtAl11}, which was motivated by the morphological analysis of a small subsample of CMASS galaxies with Hubble Space Telescope (HST) imaging. They suggest selecting galaxies with $g-i > 2.35$ produces a cleaner sample of early-type galaxies ($90\%$) that are more traditionally associated with typical LRGs. Additionally, we predict that at least a fraction of the galaxies that sit in the blue end of that colour-colour plot are also LRG progenitors, temporarily visiting the blue cloud due to small amounts of star formation. Assuming they retain their morphology (it is hard to imagine a scenario where they would not), our analysis makes quantitative predictions on the fraction of star-forming ellipticals that should be found on that part of the diagram, given the morphological mixing of the LRG sample (not currently known, to our knowledge). This result can be turned into a test of SPS models, as different sets of models will predict a different number density at those colours. We leave this exploration for future work.
%

The right-hand side panel of Fig.~\ref{fig:Vmatch_density} shows an uninterrupted trend to lower \vmatch towards fainter magnitudes. Interestingly, the slope of the \vmatch contours are almost parallel to the sliding cut in $d_\perp$ with $i-$band magnitude. This cut was designed to follow a line of constant stellar mass (Maraston et al. in prep) suggesting that the \vmatch has a clear dependence on stellar mass, as it should.

\begin{figure}
\begin{center}
\includegraphics[width=3.2in]{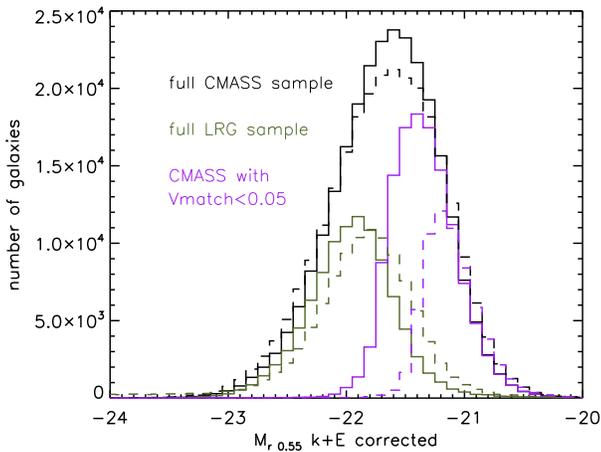}
\caption{k+e corrected absolute magnitudes for CMASS galaxies (black), LRGs (green) and the subset of CMASS galaxies that is seen in less than 5\% of the LRG volume according to our model (purple). These lie almost exclusively at the faint end, demonstrating how important the apparent magnitude cut is in the sample matching between the two surveys. Solid lines for results using the FSPS models and dashed lines for results using M11.}
\label{fig:no_LRG_magnitudes}
\end{center}
\end{figure}

A complementary way to examine the \vmatch weights is to isolate the CMASS galaxies with a small \vmatch weight - these are the CMASS galaxies that are less likely to be the progenitors of a typical LRG. Fig.~\ref{fig:no_LRG_magnitudes} shows the K+e-corrected absolute magnitude distribution of those CMASS galaxies with a $V_{match} < 0.05$, i.e. that are observed in less than 5\% of the volumes of the surveys. We clearly see these galaxies are well confined to the faint end of the CMASS population. The difference between the two models is a consequence of the steeper luminosity evolution given by K+e corrections of the M11 models - CMASS galaxies are typically brighter at LRG redshifts (when compared to a flatter luminosity evolution), and are seen through more of its volume. 

\begin{figure}
\begin{center} 
\includegraphics[width=3.2in]{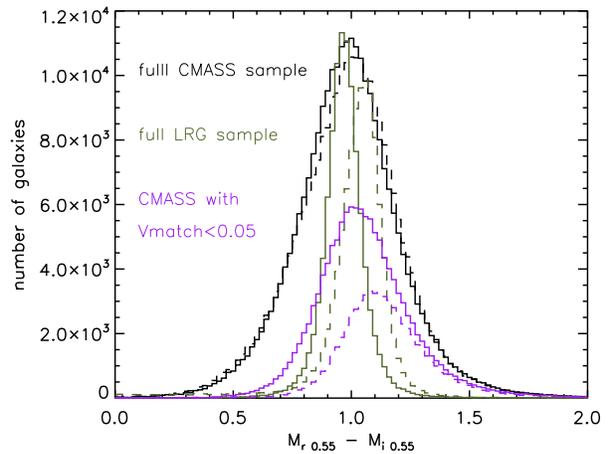}
\caption{Distribution of k+e corrected, absolute $r_{0.55}$ - $i_{0.55}$ colours for CMASS galaxies (black), LRGs (green) and the subset of CMASS galaxies (purple) that is seen in less than 5\% of the LRG volume according to our model. Solid line for results using the FSPS models and dashed line for results using M11.}
\label{fig:no_LRG_colour}
\end{center}
\end{figure}

Fig.~\ref{fig:no_LRG_colour} presents the distribution of the absolute rest-frame $r_{0.55} - i_{0.55}$ colour for the same populations as in Fig.~\ref{fig:no_LRG_magnitudes}. The bias towards losing intrinsically redder galaxies is explained by the fact that the CMASS sample is itself biased towards redder galaxies in $M_{r 0.55} - M_{i 0.55}$ at the faint end (see Section \ref{sec:samples} and Fig.~\ref{fig:colour_magnitude}) due to the $i$-band selection.

\begin{figure}
\begin{center}
\includegraphics[width=3.4in]{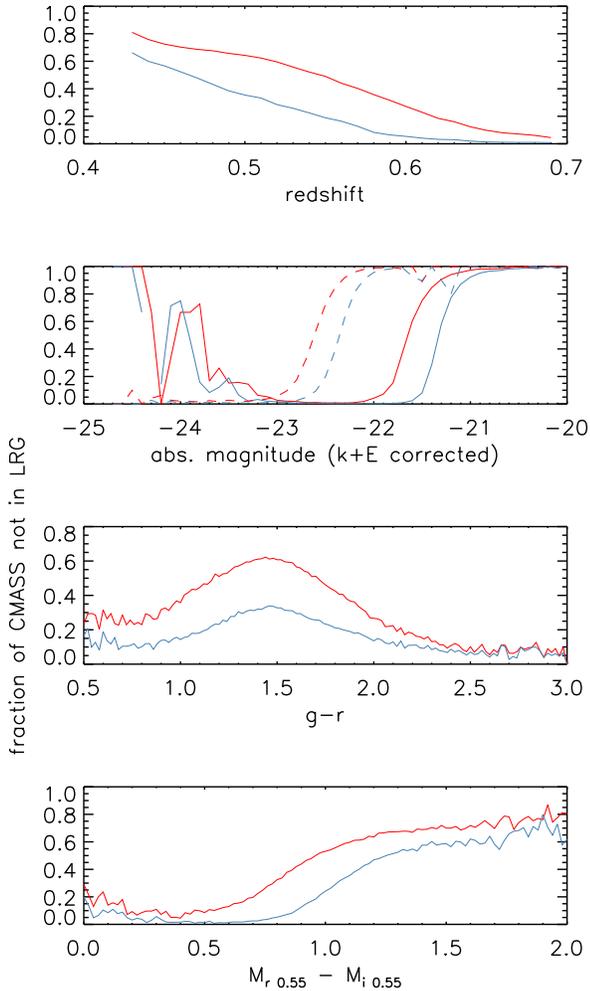}
\caption{The fraction of CMASS galaxies that is seen in less than 5\% of the LRG volume as a function of redshift (first panel), absolute magnitude (second panel - solid line for $M_{r 0.55}$ and dashed line for $M_{i 0.55}$,  observed $g-r$ colour (third panel), and k+E correct rest-frame colour $M_{r 0.55} - M_{i 0.55}$ (bottom panel). Red lines for results using the FSPS models and blue for M11.}
\label{fig:fraction_lost}
\end{center}
\end{figure}

We show the fraction of CMASS galaxies that are observed in less than 5\% of the LRG volume as a function of redshift, absolute magnitude, $g-r$ and rest-frame $M_{r 0.55} - M_{i 0.55}$ colours in Fig.~\ref{fig:fraction_lost}. Once again these figures demonstrate that magnitude is the dominant reason why these galaxies are not well matched between samples, but {\it rest-frame} colour also plays a part - see the upturn in the fraction of lost objects for bright $M_{r 0.55}$ compared to the fraction of lost objects for bright $M_{i 0.55}.$ These are the galaxies with redder $M_{r 0.55} - M_{i 0.55}$ rest-frame colours.


\section{Measuring population evolution}\label{sec:population_evolution}

In order to compute merger and luminosity growth rates, we first define the samples of CMASS galaxies and LRGs to be investigated (Section \ref{sec:sample_selection}).  Having selected matched samples, we then study the evolution of a number of quantities. In Section \ref{sec:luminosity_function} we consider luminosity functions and in Section \ref{sec:rates_of_change} the rates of change in number density, luminosity density, and typical luminosity per object.

\subsection{Sample selection}\label{sec:sample_selection}

 In each survey we take the brightest objects until we reach a given K+e corrected absolute magnitude, and we compute a \vmatch - weighted comoving number density $n$ and a \vmatch - weighted luminosity density $\ell$. We consider the following options to define the limiting magnitude in each sample, $M_{min,CMASS}$ and  $M_{min,LRG}$:

\begin{itemize}
\item A flat cut in k+E corrected absolute magnitude across the two surveys: in this case $M_{min,CMASS}$ = $M_{min,LRG}$. In general, $n_{LRG} \ne n_{CMASS}$ and $\ell_{LRG} \ne \ell_{CMASS}$.
\item A cut in K+e corrected absolute magnitude such that both samples have the same comoving number density. In this case $n_{LRG} = n_{CMASS}$ by construction, but in general $M_{min,CMASS} \ne M_{min,LRG}$, and $\ell_{LRG} \ne \ell_{CMASS}$. 
\item A cut in K+e corrected absolute magnitude such that both samples have the same comoving luminosity density. In this case $\ell_{LRG} = \ell_{CMASS}$ by construction, but in general $M_{min,CMASS} \ne M_{min,LRG}$, and $n_{LRG} \ne n_{CMASS}$.  This can be advantageous in clustering analyses that are luminosity weighted (see Section \ref{sec:clustering}).

\end{itemize}

To avoid confusion we will refer to the number and luminosity densities computed using a flat cut in absolute magnitude as $n'$ and $\ell'$.

\subsection{The luminosity function}\label{sec:luminosity_function}

With full knowledge of the completeness of the sample, we can compute luminosity functions and study their evolution. The completeness, in terms of the sample one intended to select, is primarily affected by the following well-understood effects:

\begin{enumerate}
\item targeting completeness - not all objects that pass the targeting cuts are targeted due to bright star masks, fibre collisions or other tiling issues;
\item redshift failure - not all objects with a spectrum successfully yield a redshift;
\item star/galaxy separation - galaxies that fail the star-galaxy separation in spite of being genuine galaxy targets.
\end{enumerate}

We use the targeting completeness and redshift failure corrections as described in \cite{PercivalEtAl07} for the LRGs and in \cite{RossEtAl12} for CMASS galaxies; both samples have very high spectroscopic completeness ($>97\%$). The fraction of galaxies lost to the star/galaxy separation can be estimated from commissioning data, where star-galaxy cuts are less restrictive or not included at all. This fraction is estimated to be 1\% for CMASS galaxies (Padmanabhan et al. in prep), 1\% for cut-II LRGs and $<<1\%$ for cut-I LRGs \citep{EisensteinEtAl01}. This could result in a systematic underestimate of the number density of CMASS galaxies compared to LRGs, which would at most be $\approx1\%$. In an independent analyses, \cite{MastersEtAl11} found $3\% \pm 2\%$ of CMASS targets in the COSMOS field that failed the star-galaxy cuts, in spite of being obviously galaxies when captured in high-quality HST imaging. This measurement agrees well with the numbers cited above. 

\begin{figure*}
\begin{center}
\includegraphics[width=4.5in, angle=90]{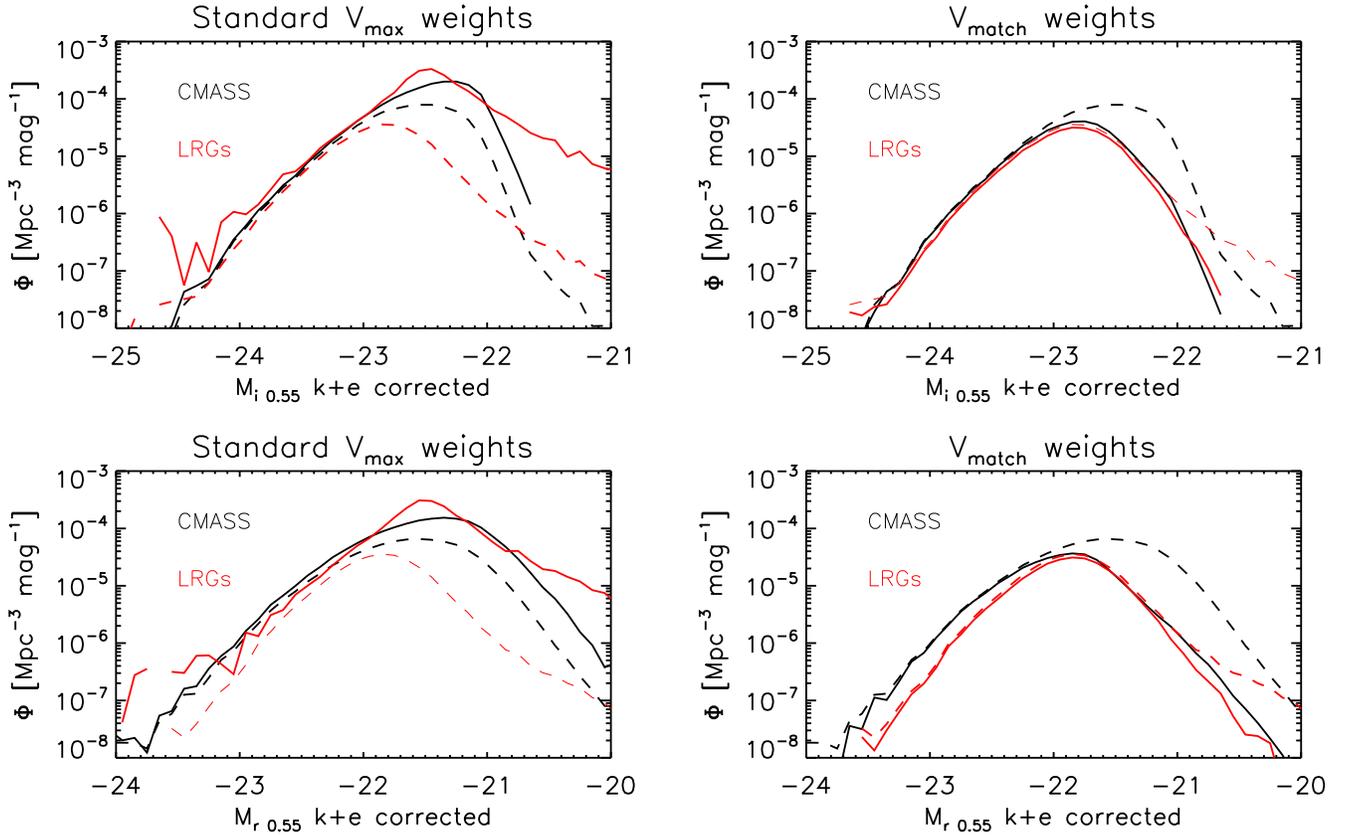}
\caption{\vmatch and \vmax weighted luminosity functions in the k+e corrected $r_{0.55}-$ and $i_{0.55}$ bands (obtained using the FSPS composite model), for the CMASS and LRG samples. The dashed lines show the un-weighted luminosity functions. The \vmax weights work by mostly up-weighting the fainter galaxies, as can be seen in the two left panels. This typically breaks down for faint galaxies. The \vmatch weight, in turn, up- and down-weights galaxies according to their relative presence on  the other survey - this can be seen in how effectively we down-weight faint galaxies in both surveys to get a luminosity function that is well matched - particularly in the $i_{0.55}-$band. Poisson errors are negligible $(\sim 1\%)$ except for the brightest or faintest half magnitudes ($1-10\%$). See text for further discussion.} 
\label{fig:luminosity_function}
\end{center}
\end{figure*}

Fig.~\ref{fig:luminosity_function} presents luminosity functions weighted by \vmatch (right), and by the standard $V/$\vmax weights (left). For reference, in both panels we show in the dashed lines the luminosity function without any completeness correction - in this case it is simply the number count of galaxies per magnitude bin, divided by the volume of each survey. We compute the luminosity function in $M_{i 0.55}$ (top) and $M_{r 0.55}$ (bottom) absolute magnitudes. Recall that the \vmatch scheme weighs each sample such that populations are matched in terms of volume, but that it does not yield true volume densities (see Section \ref{sec:weighting_scheme}). Compared to $V/$\vmax weights this {\em downweights} faint galaxies in both samples, such that the overall luminosity functions are matched. In case of zero merger evolution or contamination (and in the case of perfect modelling), our \vmatch weights fully account for changes in the {\it stellar} evolution and the two luminosity functions should therefore match. Differences can be interpreted in a number of ways: 

\begin{enumerate}
\item growth (i.e., merging); 
\item contamination: galaxies in CMASS that have identical colour and magnitudes to LRG progenitors but evolve to be something else at low redshift; 
\item resolution issues: close pairs of galaxies failing to be resolved in CMASS due to instrumental and atmospheric limitations;
\item inadequacies in the modelling - in this case, mostly in the slope of the k+E corrections. 
\end{enumerate}

It is clear that the luminosity functions of CMASS galaxies and LRGs are better matched in $M_{i 0.55}$ than in $M_{r 0.55}$. There is a larger uncertainty in the slope of the k+E corrections in the $r-$band, as that traces a region of the spectrum sensitive to small amounts of star formation at $z_c=0.55$. Small mismatches in the amount of star formation at those redshifts between our composite model and the true star-formation rate of CMASS galaxies may not be enough to down-weight them using our method, but reveal themselves in a detailed comparison such as the one we attempt here. We therefore argue that the $i-$band luminosity is more reliable for the purposes of our analysis, as it is a better tracer of overall luminosity, or stellar mass, of the galaxy. 

Differences in the {\it shape} of the luminosity function can help identify the reasons for the differences between the two samples. We present a more quantitative analysis in the next Section, where we construct three estimators to quantify differences in the amplitude and shape of the luminosity function, but first we look at the effect of using a different k+E correction model.

\begin{figure*}
\begin{center}
\includegraphics[width=4.5in, angle=90]{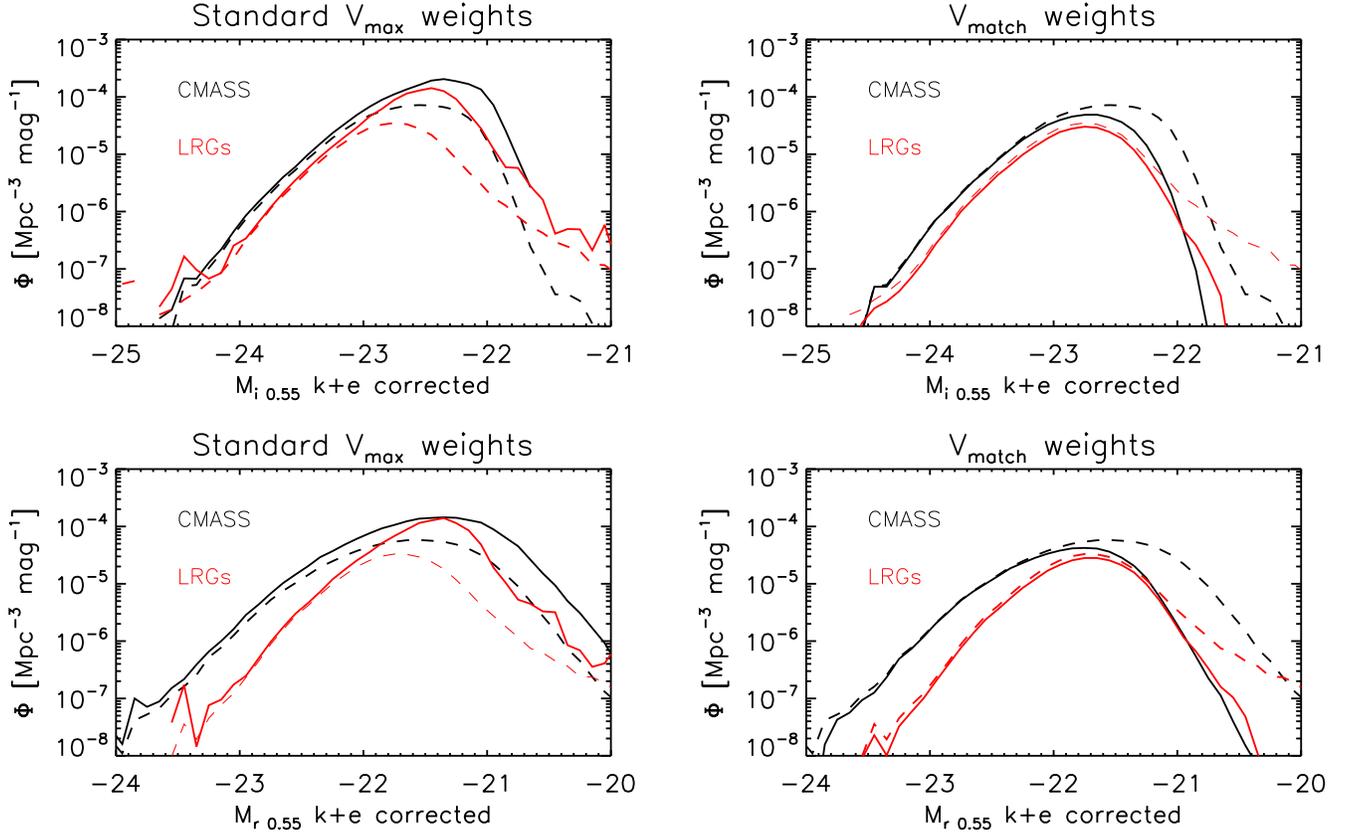}
\caption{Same as Fig.~\ref{fig:luminosity_function}, but using the absolute magnitudes computed with M11 models.}
\label{fig:luminosity_function_M10}
\end{center}
\end{figure*}

Fig.~\ref{fig:luminosity_function_M10} shows the same luminosity functions as Fig.~\ref{fig:luminosity_function}, but using the absolute k+E corrected absolute magnitudes obtained using the M11 models. The differences are substantial, especially for the LRGs. Note the differences are already apparent in the uncorrected (dashed) curves, showing the reason lies with the computation of the absolute magnitudes themselves, and not with the weighting scheme. These results are consistent with the steeper k+E correction and the magnitude distributions shown in Figs.~\ref{fig:mag_comparisons} and \ref{fig:no_LRG_magnitudes}. The effect is primarily due to the k+E corrected absolute magnitudes of the LRGs at $z_c=0.55$ - they are $\approx 0.3$ magnitudes fainter than predicted with the flatter FSPS k+E correction. The differences are larger for the LRG magnitudes simply because of our choice of $z_c$, which minimises the effect of the modelling for CMASS galaxies (see Section ~\ref{sec:ke_corrections}). 

There is an overall improvement in the matching of all \vmatch luminosity functions across the two surveys when using only red CMASS galaxies (with $g-i>2.35$) for both models. This improvement is small, of only a few per cent, and is explained by the fact that the \vmatch weights are lower for the bluer galaxies, and so they are already being down-weighted when using the full sample.

Contrasting the two weighting schemes we see that the standard $V/V_{max}$ weights up-weight galaxies at the faint end. Bright galaxies are visible in most of the survey and therefore incur a small correction. This shifts the break of the luminosity function to fainter magnitudes when compared to the uncorrected curve, but the falling in number density after that must not be trusted completely - $V/V_{max}$ weights get increasingly dominated by poisson error towards faint magnitudes (see Section \ref{sec:weighting_scheme}). This is visibly the opposite than what happens using the \vmatch weights in the opposite panels. 

\subsection{Rates of change}\label{sec:rates_of_change}

In order to understand the differences seen in Figs.~\ref{fig:luminosity_function} and \ref{fig:luminosity_function_M10} we define three estimators to quantify changes as a function of magnitude. For a pair of samples matched on luminosity density, we define a merger rate as

\begin{equation}\label{eq:r_n}
r_N = \left(1 - \frac{n_{LRG}}{n_{CMASS}}\right) \frac{1}{\Delta t},
\end{equation}

where $\Delta t$ is the time, in Gyr, between the mean redshift of the two samples (defined such that $\Delta t > 0$) . Similarly, for a pair of samples matched by number density we define a luminosity growth as

\begin{equation}\label{eq:r_ell}
r_\ell = \left( \frac{\ell_{LRG}}{\ell_{CMASS}} -1\right) \frac{1}{\Delta t}.
\end{equation}

These two rates would be exactly a merger rate and a luminosity growth in the absence of complications such as 
\begin{enumerate}
\item resolution issues: close pairs of galaxies failing to be resolved within instrumental and atmospheric limitations; 
\item contamination: galaxies in CMASS not following our composite stellar evolution model and evolving into a different region of colour and magnitude space than that of the LRGs at low-redshift;
\item loss of light to the intra-cluster medium (ICM) when a merging event occurs; and
\item a systematic offset in the computation of the absolute magnitudes as a result of the modelling.
\end{enumerate}

We investigate (i) in Section~\ref{sec:unresolved_pairs}. (ii) is an intrinsic limitation of any methodology without a full understanding of the evolution of {\em all} galaxy types. (iii) can potentially be investigated by using small-scale clustering and a halo occupation distribution type of approach, in order to estimate the fraction of satellite merging and a fraction of light lost to the ICM. We do not perform such an analysis in the present paper, but we will show in Section~\ref{sec:discussion} how, when taken together, the results we show in this and in the next Section (large scale clustering) present a picture that points strongly towards a small amount of population growth. To deal with (iv), we also define a galaxy growth rate by using our samples matched by a fixed k+E corrected absolute magnitude (see Section~\ref{sec:sample_selection}) as

\begin{equation} \label{eq:r_g}
r_g = \left(1-\frac{n'_{LRG}/\ell'_{LRG}}{n'_{CMASS}/\ell'_{CMASS}} \right) \frac{1}{\Delta t}
\end{equation}

$r_g$ would match the merger rate even in the presence of contaminants (assuming the luminosity function of the contaminants was the same as the luminosity function of the CMASS galaxies). More generally, it can be interpreted as a rate of change of luminosity per single object across the two surveys. Whereas $r_N$ and $r_\ell$ are dominated by the relative amplitude of the luminosity function between the two redshifts, $r_g$ tells us about differences in the shape. 

\subsubsection{Results}\label{sec:population_growth}

\begin{figure}
\begin{center}
\includegraphics[width=3.2in]{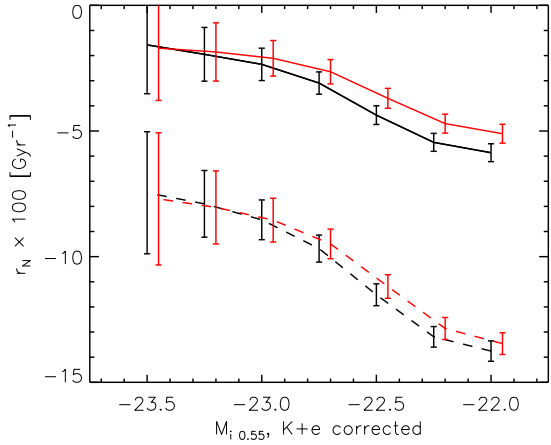}
\caption{The merger rate, per Gyr, computed as per equation \ref{eq:r_n} as a function of the magnitude of the faintest LRG in the sample. The black lines shows $r_N \times 100$ for the full sample, and the red line for galaxies with $g-i>2.35$. The results obtained from using M11 models (dashed lines) show the same slope with magnitude as the results using FSPS models (solid lines), but are a {\em factor of two to three} lower. Poisson errors shown.}
\label{fig:merger_rates}
\end{center}
\end{figure}

\begin{figure}
\begin{center}
\includegraphics[width=3.2in]{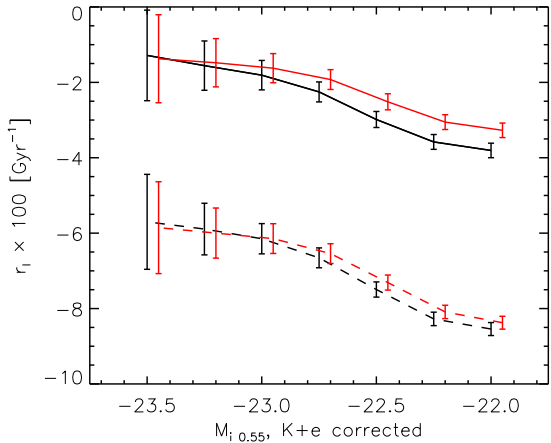}
\caption{The luminosity growth, per Gyr, computed as per equation \ref{eq:r_ell} as a function of the magnitude of the faintest LRG in the sample. The black lines shows $r_\ell \times 100$ for the full sample, and the red line for galaxies with $g-i>2.35$. The results obtained from using M11 models (dashed lines) show the same slope with magnitude as the results using FSPS models (solid lines), but are a {\em factor of two to three} lower. Poisson errors shown.} 
\label{fig:luminosity_growth}
\end{center}
\end{figure}

We compute $r_N$ and $r_\ell$ as a function of $M_{i 0.55}$ (the magnitude of the faintest LRG in the sample, which was used to compute the matched samples - see Section \ref{sec:sample_selection}), which are shown in Figs.~\ref{fig:merger_rates} and \ref{fig:luminosity_growth}. Our most inclusive samples (i.e., where $M_{i 0.55}=-22$) include $\approx 95\%$ of the LRGs and $\approx 40\%$ of CMASS galaxies, and have large stellar masses with $\log_{10} M/M_{\odot} \gtrsim 11.2$ (Maraston et al. 2012, in prep).

$r_N$ is negative for all magnitudes, although it tends to zero towards brighter magnitudes. This implies that, for the same integrated luminosity density, there are {\it more} LRG galaxies per comoving volume than there are CMASS galaxies. I.e., CMASS galaxies appear to be {\it brighter} than LRGs in the $i_{0.55}$ band. This is expected from our analysis of the luminosity functions of Figs.~\ref{fig:luminosity_function} and \ref{fig:luminosity_function_M10}. We emphasise that if this brightening was due simply to the stellar evolution, and in the absence of other complications, then our model and \vmatch weights would account for it.

$r_\ell$ naturally tells a similar tale - for the same comoving number density, LRGs are hold less luminosity than CMASS galaxies. Removing galaxies with observed colour $g-i<2.35$ reduces this number by $\lesssim 1\%$ at the faintest magnitudes, but a 5\% discrepancy remains, even for the reddest galaxies in the CMASS sample. As is obvious from the luminosity functions in Figs.~\ref{fig:luminosity_function} and \ref{fig:luminosity_function_M10}, these rates are heavily dependent on the slope of the k+E corrections. Results using the M11 models are identical in shape, but are lower by a factor of two to three. I.e. - the uncertainty in the modelling of the k+E corrections can potentially overwhelm these statistics. We return to this at the end of this section.  One point of interest is how the \vmatch $M_{i 0.55}$ CMASS luminosity function seems offset from that of the LRGs by an almost constant factor as a function of magnitude for both FSPS and M11 - this is likely a result of a k+E correction slope that is too steep. 

\begin{figure*}
\begin{center}
\includegraphics[angle=90, width=7in]{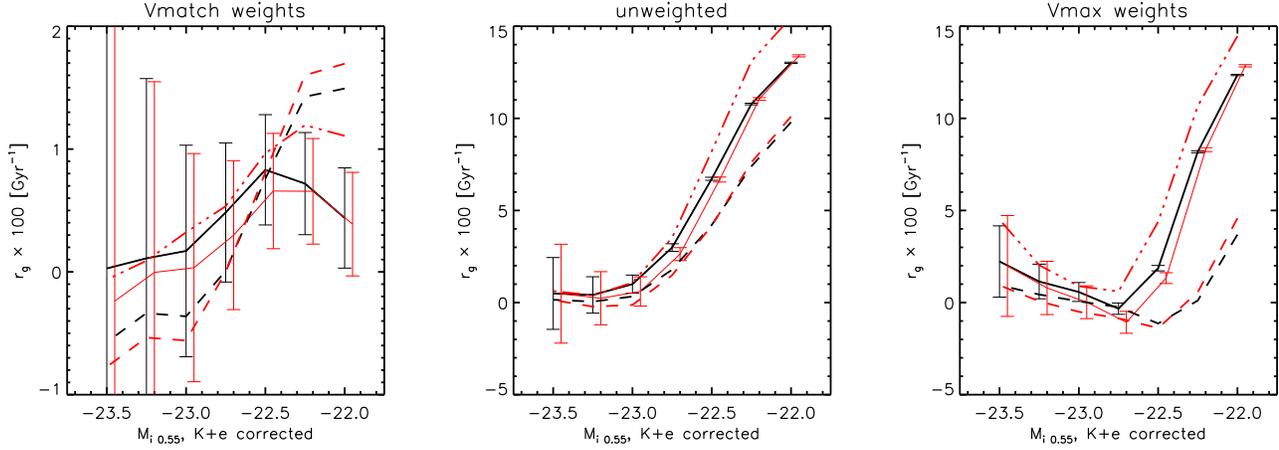}
\caption{The change in average light per luminosity, per Gyr, computed as per equation \ref{eq:r_g}  as a function of faintest galaxy in the sample. The three panels show the different weighting schemes used when computing number and luminosity densities: \vmatch on the left, unweighted in the centre, and \vmax on the right. The black lines shows $r_g \times 100$ for the full sample, and the red line for galaxies with $g-i>2.35$. When weighted by \vmatch, $r_g$ shows evidence for a slowly evolving population using both stellar population models (M11 in dashed lines, FSPS in the solid lines; we also show the purely passive model of \protect\citealt{MarastonEtAl09} in the dot-dashed red line - see Section \ref{sec:passive_model} for details).  The trend in the middle panel is dominated by incompleteness issues in the LRG sample, which are severe for $M_{i 0.55} > -23$ (see Figs.~\ref{fig:luminosity_function} and \ref{fig:luminosity_function_M10}). $V/$\vmax weights (right) result in a low $r_g$ down to lower magnitudes than $V/$\vmax, but it rises a steeply with decreasing luminosity beyond that. This could be a result of an inadequate completeness correction, or increased merging rate at these luminosities. In any case, this comparison demonstrates clearly that the way in which the \vmatch weights balance the two samples at low luminosities results in a well matched sample in terms of comoving densities and average luminosity per galaxy - as is our goal. Poisson errors are shown for one of the sets of models only for clarity - they are identical for the other set. See text for further discussion.} 
\label{fig:r_g}
\end{center}
\end{figure*}

To help understand the observed evolution, we examine the rate of change in weighted luminosity per object, or $r_g$ as given by equation (\ref{eq:r_g}), which we show on the left-most panel of Fig.~\ref{fig:r_g}. Recall that, for this statistic, we select galaxy samples based on a fixed k+E corrected absolute magnitude. Using either SPS model, $r_g$ is between $-1\%$ (at the bright end) and $2\%$ (at the faint end). A steeper evolution seen with M11 is now clear, and it indicates that the typical luminosity per galaxy increases between the two surveys, especially at the faint end. A similar trend is seen using FSPS models, but it is less significant. Processes like merging would act to change the shape of the luminosity function, according to the fraction and magnitude of the merging galaxies. However, that is not what is observed in the $M_{i 0.55}$ luminosity function with {\it either} set of models. In other words, the fact that we observe a small value of $r_g$ is {\it support for a slowly evolving weighted luminosity per galaxy} between the two surveys. Note that the sign is positive - i.e. $\frac{n'_{LRG}/\ell'_{LRG}}{n'_{CMASS}/\ell'_{CMASS}} < 1$, or in other words there is on average more luminosity per galaxy in the LRG sample. This is now consistent with a small amount of luminosity growth through merging.

For comparison, we also show $r_g$ computed using unweighted number and luminosity densities, or using \vmax-corrected densities. In the unweighted case, we see a much steeper trend in inferred merger rate with luminosity. This trend is dominated by incompleteness issues within the LRG sample, which becomes serious at around $M_{i 0.55} = -23$, as can be seen in the dashed lines of Figs.~\ref{fig:luminosity_function} and \ref{fig:luminosity_function_M10}. A $V/$\vmax weight results in a lower inferred merger rate down to lower magnitudes ($M_{i 0.55} = -22.5$), but shows a steep trend of increasing $r_g$ with decreasing luminosity beyond that. It is difficult to assess whether this effect is due to $V/$\vmax being insufficient to fully correct for completeness or whether it is due to a steeper merging rate at those luminosities (which in turn are down-weighted using the \vmatch approach). In any case, this comparison demonstrates quite clearly that the way in which the \vmatch weights balance the two samples at low luminosities results in a well matched sample in terms of comoving densities and average luminosity per galaxy. 

To summarise: we have a complicated scenario: $r_N$ and $r_\ell$ only reflect a true merger rate or luminosity growth in the absence of contamination or unresolved pairs, and a true contamination/unresolved pairs fraction in the absence of merging. These two quantities are also sensitive to a change in the slope of k+E corrections as they rely on matching samples by luminosity and number density. They show a significant excess of luminosity in CMASS, with respect to what we should expect from LRGs. $r_g$, measuring the change in the average luminosity per object, is less sensitive both to the slope of the k+E correction and to contaminants (provided they have a similar luminosity than the galaxies of interest). This quantity shows a modest evolution between the two surveys ($<2\%$) for both stellar population synthesis (SPS) models. 

We take this investigation further by seeing whether unresolved pairs in CMASS could explain the excess of luminosity implied by $r_N$ and $r_\ell$ alone.

\subsection{Unresolved pairs}\label{sec:unresolved_pairs}

\begin{figure}
\begin{center}
\includegraphics[width=3.2in]{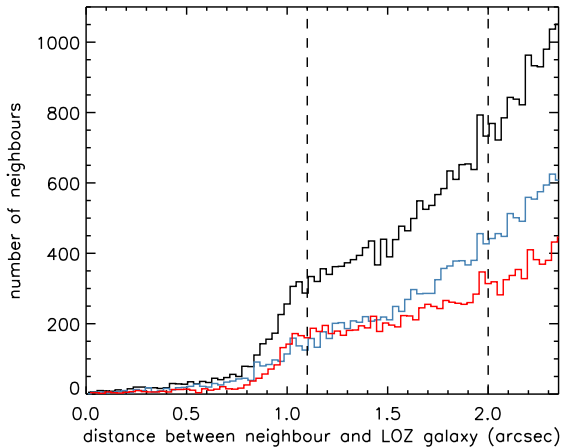}
\caption{Number of LOZ galaxies with a photometric pair as a function of its distance. The blue line shows the distances distribution for photometric pairs classified as stars, and the red line shows the distances for photometric pairs classified as galaxies. The black like is the sum of the two. The vertical dashed lines are representative of the seeing discs at $z=0.3 (1.2")$ and the angular size of the 1.2" seeing disc at $z=0.6$ redshifted to $z=0.3 (2")$. The clear drop off in the number of pairs at distances smaller than roughly $1"$ is due to the fact that we cannot resolve pairs closer than the seeing disc.} 
\label{fig:d_neighbours}
\end{center}
\end{figure}

We investigate this issue by looking at pairs of LRGs with another object (photometrically classified as a  galaxy), within a $\approx 2"$ separation - the angular size subtended by same the physical distance at $z=0.3$ that corresponds to 1.2" at $z=0.55$. In other words, we find all LRGs with a close companion such that they would be likely unresolved due to seeing (taken to be typically 1.2") at CMASS redshifts. In order to increase our statistics, and to allow us to investigate this issue to fainter magnitudes, we perform this analysis in the LOZ sample. The LOZ targeting is very similar to that of the LRGs in terms of colour, but targets fainter galaxies (see Section~\ref{sec:data}). We use the full photometric sample as this sample is very pure, with stellar contamination at less that $2\%$ (Padmanabhan et al. in prep). We apply the cuts described in Section~\ref{sec:data} on DR8 photometry \citep{AiharaEtAl11}, resulting in approximately 1 million targets. Of these, only $\approx$ 15,000 (30,000), or roughly $1.3\%$  $(2.4\%)$ have a pair between $1.2"$ and $2"$ $(2.4")$. Approximately half of these close neighbours are photometrically classified as galaxies and half are photometrically classified as stars. We also note that the foreground volume of a $2"$ arcsec disc at $z=0.3$ is roughly one third of the foreground volume of a $1.2"$ arcsec disc at $z=0.55$. So assuming a constant number density of foreground objects, we should multiply our estimate of the multiple fraction due to chance alignment by a factor of 3. We have no way to estimate how many of the the close pairs are chance alignments and how many are physically associated pairs. If we assume that all pairs are chance alignments we reach an estimate on the number of unresolved targets at CMASS redshifts of $\approx 2\%$. We show the distance profile of these pairs in Fig.~\ref{fig:d_neighbours}.

\begin{figure}
\begin{center}
\includegraphics[width=3.7in]{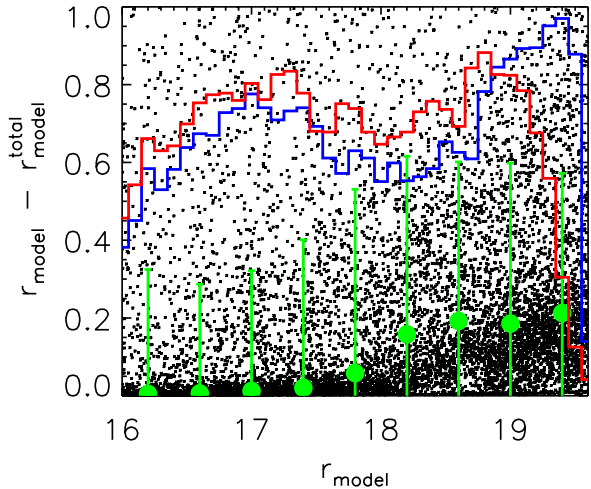}
\caption{The excess magnitude in $r-$band introduced by potentially unresolved pairs from neighbours photometrically classified as galaxies, as a function of $r-$band magnitude (black dots, with the median in green). The histograms show the distribution of $r_{mod}$ before (blue) and after (red) adding the flux of the close neighbour (histograms are not normalised to the y-axis, but share a common normalisation). } 
\label{fig:mag_extra}
\end{center}
\end{figure}

We can use our estimate of unresolved multiples to calculate the additional flux brought into the CMASS sample from potentially unresolved neighbours. Fig.~\ref{fig:mag_extra} shows the excess in the $r-$band as a function of $r_{mod}$ from pairs classified as galaxies (black dots, with the median in green). We over plot the unnormalised distributions of $r_{model}$ before (blue) and after (red) adding the flux of the close pair. It is clear that the effect can be quite dramatic (approximately $0.2$ magnitudes) for objects fainter than $r_{mod}\approx 18$ magnitudes. It is worth pointing out that in spite having of smaller statistics, if we repeat the above analysis around LRG targets (as opposed to LOZ galaxies) we get perfectly consistent results. 

Integrated over all galaxies with close neighbours, the total flux brought into the sample by neighbours photometrically classified as galaxies is roughly 7.5\% of the $r-$band flux of the LRGs. However, this is only happening to roughly 2\% of the CMASS sample according to our more generous estimate, and is therefore too small to explain the observed excess in luminosity in CMASS, compared to what is expected from LRGs {\it if} we attribute this excess of luminosity to unresolved targets. Reversing the question, to explain the excess in luminosity that we observe in the case of FSPS models at the faintest end (a 4\% excess in luminosity integrated over the sample; see Fig.~\ref{fig:luminosity_growth}), we require that over 50\% of the CMASS galaxies are in fact unresolved targets due to chance alignments (this number would have to increase by approximately a factor of two to explain the excess in luminosity inferred using the M11 models). This is 25 times larger than the fraction estimated by our analyses of close pairs in LOZ, suggesting that the slope of k+E corrections or contamination, rather than unresolved targets, is the mostly the source for the trends seen in Figs.~\ref{fig:merger_rates} and \ref{fig:luminosity_growth}, and explains why we see only a small evolution in $r_g$ in Fig.~\ref{fig:r_g}.

\cite{MastersEtAl11} identified a significant number of unresolved targets in CMASS by looking at HST COSMOS data of a small sub-sample of CMASS galaxies. They show that  $\approx 21 \pm 4\%$ of CMASS galaxies are in fact unresolved pairs, of which approximately half are estimated to be a result of chance alignment, and half physically connected pairs (i.e. satellites in the same dark matter halo). Note that we are not interested in any unresolved CMASS objects that are also unresolved at low redshift, as any such close neighbours will have their flux accounted for in our estimates of the LRGs luminosity, and therefore would not contribute towards the discrepancy shown in Figs.~\ref{fig:merger_rates} and \ref{fig:luminosity_growth}. Without information on the radial distribution of the unresolved CMASS targets analysed by Masters et al., a direct comparison is not particularly insightful.
 
Our analysis shows clearly that unresolved targets cannot account for the excess in luminosity observed in CMASS galaxies. This remains true even allowing for the fraction of unresolved pairs measured by Masters et al. Once again we emphasise that the slope of k+E corrections, rather than unresolved targets, is most likely the reason for the trends seen in Figs.~\ref{fig:merger_rates} and \ref{fig:luminosity_growth}, and explains why we see only a small evolution in $r_g$ in Fig.~\ref{fig:r_g}.

\subsection{Beyond the composite model}\label{sec:beyond_composite}

In Section \ref{sec:composite_model} we introduced the composite model as the best estimate of the overall average colour and magnitude evolution of the full LRG sample. In this Section we take the opportunity to briefly consider two additional approaches to modelling the colours and magnitudes of LRGs. Firstly, we consider a purely passive stellar model and secondly we consider using all individual LRG models instead of averaging them into a single composite stellar prescription. 

\subsubsection{A passive model}\label{sec:passive_model}

We take the fully passive stellar model of \cite{MarastonEtAl09} and assume it applies to every LRG and to every red CMASS galaxy (with $g-i<2.35$). CMASS galaxies bluer than this cut show signs of star formation (Thomas et al. in prep) and a significantly distinct morphological mix with a large fraction of late-type galaxies \citep{MastersEtAl11};  a purely passive model is simply not a correct description of the bluer CMASS galaxies. The \cite{MarastonEtAl09} passive model is based on a single burst of star formation at $z\sim5$, with solar metallicity and an additional component of 3\% (by mass) of metal-poor old stars. As this model was fitted to 2dF SDSS LRG and Quasar (2SLAQ) data \citep{CannonEtAl06}, with a significantly redder selection function, it does not fit the CMASS data as well as our composite model (see Fig.~\ref{fig:composite_model}).

We proceed in exactly the same way as for the composite model, and we show the resulting evolution of $r_g$ with $M_{i0.55}$ as the dot-dashed red line in Fig.~\ref{fig:r_g}. The new results  agree particular well with the results using the FSPS composite model, as expected given Fig.~\ref{fig:ke_corrections}: the k+E correction for a purely passive model follows closely that of the FSPS composite model k+E corrections. This shows that the differences in the modelling stem mostly from differences in the assumed star-formation histories, but note that these in turn are driven by the different stellar evolution tracks assumed in each set of stellar population synthesis models. Once we weight by \vmatch all models give consistent results. The differences are more significant without the \vmatch weight, and the inferred merger rates are then larger with a purely passive model. 

\subsubsection{Using 124 stellar evolution models}\label{sec:allstacks}

Whereas the composite model captures the average stellar evolution of the LRGs, there is a significant amount of scatter around this average especially in the case of the M11 models. We attempt to use the full range of individual fits to LRGs of different colour, redshift and luminosity as follows. To each LRG we assign the correct model according to its $r-i$ colour, luminosity and redshift. To each CMASS galaxy we assign the closest model in terms of colour: we take the predicted $g-r$ and $r-i$ colours of the 124 models at the redshift of each CMASS galaxy, and we assign to that galaxy the model that sits closest. Note that as photometric scatter is larger than the typical distance between different models this is an intrinsically noisy process.

\begin{figure}
\begin{center}
\includegraphics[width=3.5in]{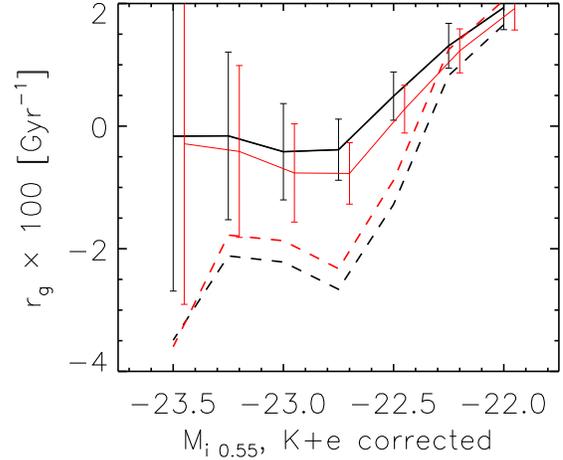}
\caption{The change in average light per luminosity, per Gyr, computed as per equation \ref{eq:r_g}  as a function of faintest galaxy in the sample, using 124 models of stellar evolution as described in Section \ref{sec:allstacks}. The black lines shows $r_g \times 100$ for the full sample, and the red line for galaxies with $g-i>2.35$. M11 results are shown in dashed lines, and FSPS results in solid lines. These results should be compared to the left-most panel of Fig.~\ref{fig:r_g}, obtained using the composite model. The results are similar in the FSPS case, but steeper for the M11 models. The differences are not significant given the errors. }
\label{fig:r_g_allstacks}
\end{center}
\end{figure}

As explained in Section \ref{sec:composite_model} each stellar evolution model has a different scope in redshift. We therefore must change the definition of our weights to allow for this fact: we use only the volume probed by each stellar evolution model to define $V_{\rm{LRG}}$, $V_{\rm{max}}^{LRG}$, $V_{CMASS}$ and $V_{\rm{max}}^{CMASS}$ in Eqs. (\ref{eq:wa}) and (\ref{eq:wb}). This is equivalent to splitting into 124 pairs of surveys, each with a different predicted evolution and redshift range, and considering the result on the combination. We show the resulting evolution of $r_g$ in Fig.~\ref{fig:r_g_allstacks}.

 When compared to the values obtained with the composite model we see a larger difference in $r_g$ in the case of M11 models; this is expected given the larger scatter around this composite model. Given how noisy the process of assigning a model to each CMASS galaxy can be, it is hard to interpret this difference, which is not in any case significant. Whereas it is obviously desirable to include as much information as possible on the stellar evolution of the LRGs, photometric scatter makes this method unreliable. One potential improvement over the composite model is to construct two or more average models, for sufficiently distinct areas of colour-colour or colour-magnitude space. We leave such explorations for future work. For the moment  we emphasise that the composite model we describe in Section \ref{sec:composite_model} is the most suitable choice for the current analysis.

\section{Large-scale clustering}\label{sec:clustering}

The evolution of the large scale clustering has the potential to help us interpret the results of the previous sections by constraining the merging history of the sample. The evolution of the large-scale linear bias has a well-defined evolution, given by \cite{Fry96}, for pure passive evolution (i.e., no mergers). This gives us the opportunity to check whether our weighted samples are consistent with a small amount of merging and contamination, as suggested by Fig.\ref{fig:r_g}.

We will follow \cite{TojeiroEtAl10} and weigh galaxies by their luminosity. The advantage is that any merging happening amongst CMASS galaxies will not contribute towards a deviation from the Fry et al. model, provided that no significant loss of light happens to the ICM. Note that even if this loss is significant, it is still preferential to match samples by luminosity density and weight by galaxy luminosity - see Section \ref{sec:discussion}.

\subsection{Measuring and modelling the correlation function}

The two-point correlation function, $\xi(r)$ measures the excess probability, $dP(r)$ of finding a pair of galaxies at a given distance $r$, compared to a purely random distribution:

\begin{equation}
dP(r)  = n[1+\xi(r)] dV 
\end{equation}
In practice, we count pairs of galaxies in bins of $r$ and $\mu$, and use the \cite{LandySzalay93} estimator as

\begin{equation}\label{eq:xi_estimator}
\hat{\xi_\ell}(r) = \frac{\sum_{\mu} DD(r,\mu) - 2DR(r,\mu) + RR(r, \mu) }{\sum_\mu RR(r,\mu)} P_\ell
\end{equation}
where $DD$, $DR$ and $RR$ are normalised galaxy-galaxy, galaxy-random and random-random pair counts in bins of $r$ and $\mu$ respectively ($\mu$ is the cosine of the angle between a galaxy pair and the line of sight). We use a random catalogue with the same angular mask as the data catalogue, and with an $n(z)$ matched to that of the data. To avoid contributions from shot noise from the random pair counts, we use random catalogues with 10 times the number density of the data. 

Setting $\ell = 0$ in equation~(\ref{eq:xi_estimator}) gives us the monopole of the correlation function, as defined in \cite{Hamilton92} - this is the excess of finding a pair of galaxies at given distance $r$ averaged over pairs observed at all angles with respect to the line of sight. 
The quadropole, or $\ell=2$ contains the next order of information, by effectively comparing the power along and across lines of sight. $\xi_0$ and $\xi_2$ are both affected by  redshift-space distortions and enhanced clustering along the line-of-sight, which we model. Even though the passive model of \cite{Fry96} constrains only the spherically averaged power, or $\xi_0$, we fit our data to models of $\xi_0$ and $\xi_2$, as this improves our signal. 

We model the isotropic, $\mu$-averaged  correlation function $\xi(r)$ as in \cite{SamushiaEtAl12}. A non-trivial survey geometry imprints a non-uniform distribution of pairs in $\mu$ on the data as not all galaxy-pair configurations are allowed by the window function. We correct for this effect as in \cite{SamushiaEtAl12}, by weighting each galaxy pair such that the weighted distribution of pairs in $\mu$ corresponds to that expected in the absence of a window function.


\subsection{Fitting the correlation function}

To increase our resolution with redshift, we split each of the CMASS and LRG slices into two, giving a total of 4 luminosity-matched slices centred at $z=$0.3, 0.4, 0.5 and 0.6. For each of the slices we compute $\hat{\xi}_0(r)$ and $\hat{\xi}_2(r)$ according to Equation~\ref{eq:xi_estimator}, and we use a simple 2-dimensional $\chi^2$ minimisation in order to find the best fitting scale-invariant amplitudes. 

We estimate the errors and their covariance by using mock simulations. We use mock catalogues constructed using the Large Suite of Dark Matter Simulations (LasDamas, McBride et al. in prep) in order to construct 80 independent realisations of $\hat{\xi}_0$ and $\hat{\xi}_2$ for the first two redshift slices (we sub-sample each mock in order to reproduce the $n(z)$ in each redshift slice). For the last two redshift slices, we use 600 perturbation-theory halo mocks of \cite{ManeraEtAl12}, and follow the same procedure. 
To ensure a stable inversion of the covariance matrix, and to increase our signal-to-noise in each bin, we re-bin the correlation functions to 11 bins in comoving distance, logarithmically spaced between 30 and 200 Mpc/h. This results in a total of 22 measurements to be fitted by two parameters, totalling 20 degrees of freedom.


\subsection{Large-scale bias evolution}
\begin{figure*}
\begin{center}
\includegraphics[width=3.4in]{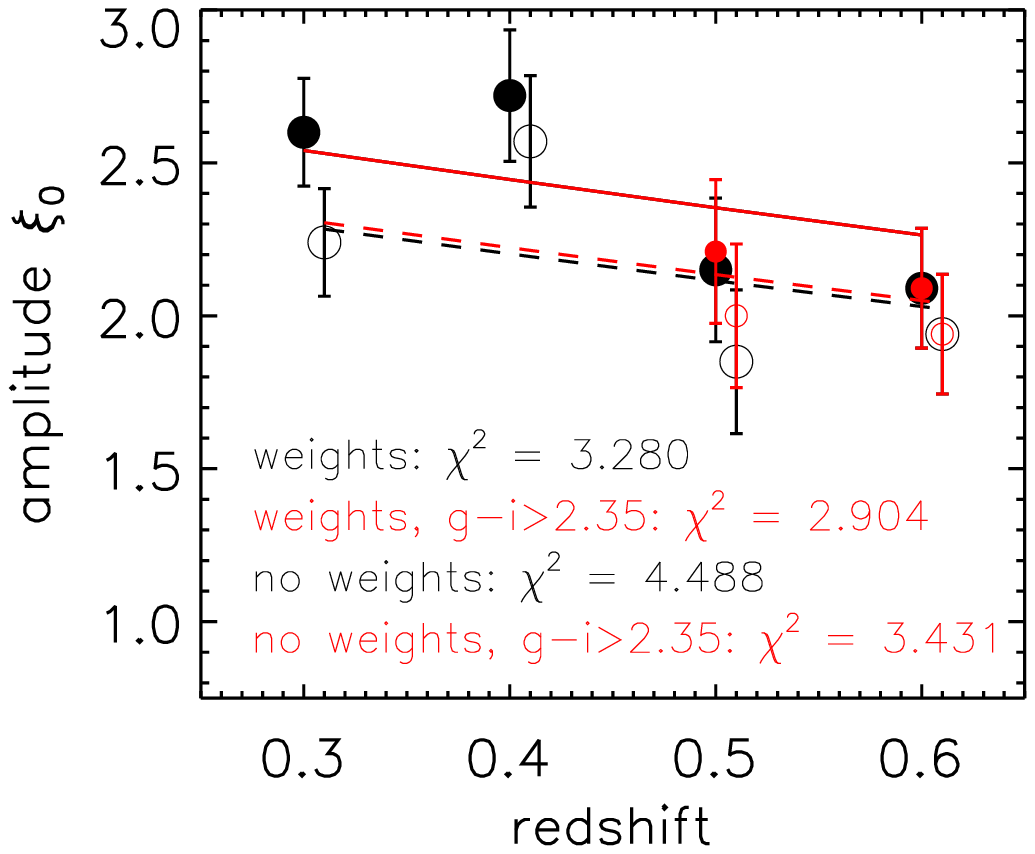}
\includegraphics[width=3.4in]{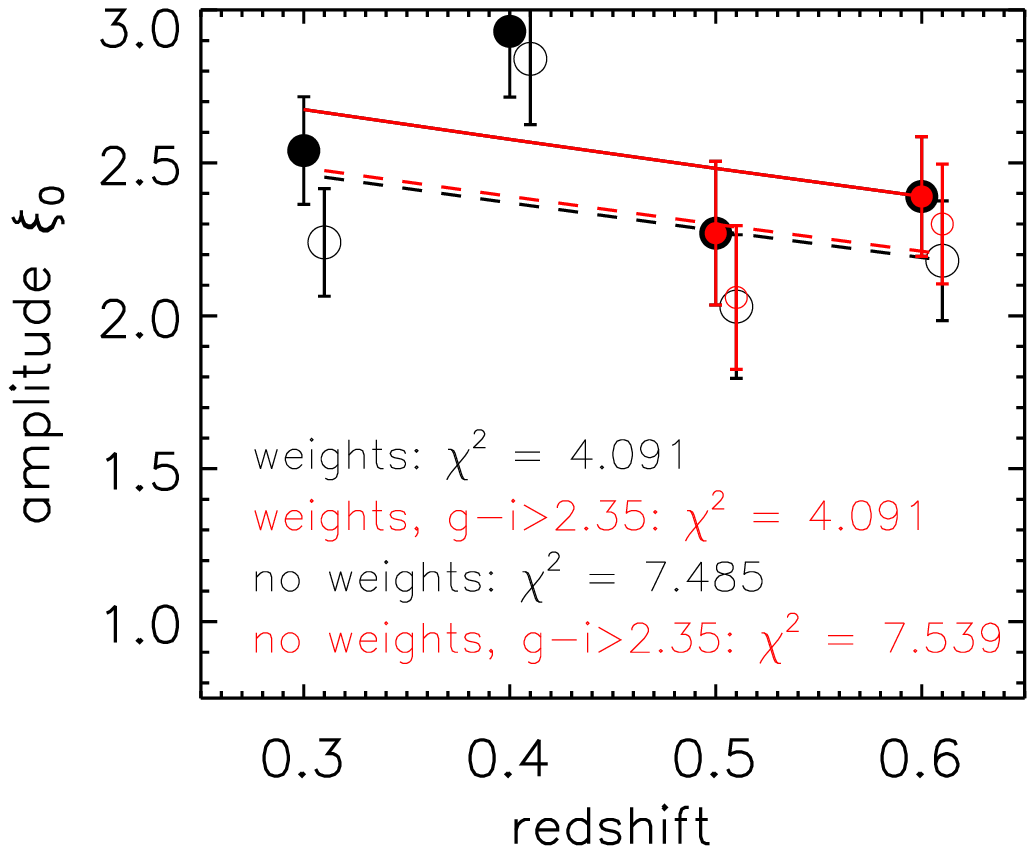}

\caption{The amplitude of the luminosity and \vmatch-weighted large-scale amplitude of $\xi_0$ computed as a function of redshift (solid black dots). The two lowest  redshift points lie exclusively within the LRG sample, and the two highest redshift points exclusively in the CMASS sample. The open dots show the amplitude of the un-weighted large-scale power, fitted to the same scales. The red dots (open and filled) show the amplitude of $\xi_0$ when selecting only CMASS galaxies with $g-i>2.35$. The lines show the best fit passive model of \protect\cite{Fry96} (Equation \ref{eq:amplitude}), obtained by assuming LCDM and fitting for $b(z_0)$ - the solid line is a fit to the filled (weighted) points, and the dashed line is a fit to the open (not weighted) points. We give the minimum values of $\chi^2$ for each case. Left: FSPS models; right: M11 models. }
%
\label{fig:xi0_ev}
\end{center}
\end{figure*}

The evolution of the amplitude of the monopole can be seen in the filled circles of Fig.~\ref{fig:xi0_ev}, with error bars derived from the fits to all of the mocks, using the covariance matrices described in the previous section.

To check whether our results are consistent with the \cite{Fry96} evolution, we model the redshift evolution of the amplitude of the monopole, $A_0(z)$, 
as \citep{Hamilton92}:

\begin{equation}\label{eq:amplitude}
A_0(z) = \left( b^2(z) + \frac{2}{3}f(z)b(z) + \frac{1}{5} f^2(z) \right) \sigma_8^2(z)
\end{equation}
with $\sigma_8(z) = \sigma_8(0)D(z)/D(0)$ and 
\begin{equation} \label{eq:fry96}
b(z) = [b(z_0) - 1] \frac{D(0)}{D(z_0)} + 1
\end{equation}
$f$ is the logarithmic derivative of the linear growth factor $D(z)$ with expansion, $f\equiv d\log D(z)/d\log a$. For simplicity we assume a LCDM cosmology and $f(z) = \Omega_m^{\gamma(z)}$, with $\gamma(z) = 0.557 - 0.02z$ (\citealt{PolarskiEtAl08}). 

We take $z_0 = 0.3$ and use a simple $\chi^2$ minimisation to fit the model of Equation \ref{eq:amplitude} to our four data points. We perform this analysis eight times:
\begin{enumerate}
\item by weighting the galaxies by their luminosity and \vmatch;
\item by not weighting the galaxies;
\item by weighting the galaxies by their luminosity and \vmatch and applying a $g-i>2.35$ cut; and
\item by not weighting the galaxies and applying a $g-i>2.35$ cut
\end{enumerate}
with each of the two stellar population models. The $\chi^2$ values of our fits can be seen in Fig.~\ref{fig:xi0_ev}. For both the FSPS and M11 results, the best-fit comes from when the data is weighted by \vmatch and by luminosity, and when we use only galaxies redder than $g-i=2.35$. This result is a good indication that our weights are performing as expected, and that there is less evolution towards the red end of the galaxy population, as expected. Weighting by \vmatch and luminosity makes a larger difference to the best-fitting $\chi^2$ than cutting the galaxies in colour - this is because the \vmatch weights effectively down-weight blue galaxies most effectively. The weights increase the overall amplitude of $\xi_0$ simply because we are up-weighting the most luminous objects and these are more biased (e.g. \citealt{ZehaviEtAl05b,ZehaviEtAl11}). FSPS models give a formally better fit than M11 models, but note that in the case of weighted red galaxies, both models give acceptable fits to the passive model. This result is a welcome confirmation of our interpretation of Fig.~\ref{fig:r_g}. In summary, our weights and sample matching yield a sample of galaxies that is consistent with dynamical passive evolution. Moreover, this is robust to the set of stellar population models used to create the samples and compute the weights.

\section{Summary, discussion and conclusions}\label{sec:discussion}

In this paper we present a joint analysis of SDSS-I/II LRGs and BOSS CMASS galaxies, with the aim of identifying and characterising a coeval population of galaxies spanning a redshift range between $0.23$ and $0.7$. We are motivated from the desire to select a population of galaxies that is evolving passively in a dynamical sense (i.e. no mergers) as closely as possible, as the large-scale bias evolution of such a population can be understood analytically and yield significant gains in cosmological analysis of large-scale structure. 

We focused on the progenitors of LRGs, as massive red galaxies are prime candidates for such a population. As the targeting selection in CMASS is significantly wider both in colour and magnitude (Section~\ref{sec:data}), we developed a set of weights that optimally matches galaxies in terms of their stellar evolution (Section \ref{sec:weighting_scheme}). To do so, we relied on the fossil record of LRGs from which we extracted a stellar evolution model (Fig.~\ref{fig:composite_model}). We then used this model to identify the most likely progenitors of LRGs amongst CMASS galaxies (Section~\ref{sec:progenitors}). Finally we developed a number of estimators to attempt to characterise and quantify population evolution between the two surveys (Sections \ref{sec:population_evolution} and \ref{sec:clustering}).

We find that CMASS galaxies are an extension of LRGs by being both intrinsically fainter and having a wider range in observed and in rest-frame colour. Fig.~\ref{fig:fraction_lost} shows the fraction of CMASS galaxies that are not expected to evolve into LRGs as a function of redshift, absolute magnitude, observed $g-r$ colour and rest-frame $M_{r 0.55} - M_{i 0.55}$ k+E corrected colour. We find a steep dependences in absolute magnitude and rest-frame colours, confirming that CMASS galaxies are broader in terms of {\it intrinsic} properties. 

Our analysis of weighted number and luminosity densities using the $r_N$ and $r_\ell$ estimators (Figs.~\ref{fig:luminosity_growth} and \ref{fig:merger_rates}) points towards a scenario where the CMASS sample is typically brighter expected from the LRG progenitors. To investigate this issue further we considered the potential contamination from unresolved targets - i.e. CMASS targets that are targeted as a single object, but are in fact unresolved pairs of stars of galaxies (Section \ref{sec:unresolved_pairs}). By examining close pairs of LOZ galaxies (such that they would be unresolved at CMASS redshifts) we estimate that the extra luminosity from these pairs would be too small ($<1\%$ in the $r^{0.55}-$band) to explain the excess in luminosity we see in CMASS, even in our most generous scenario. The most likely reason for the remaining differences is uncertainty in the slope of the k+E corrections (which affect the absolute magnitudes mostly of the LRGs), or contaminants - objects in CMASS that share a region of colour-magnitude space with LRG progenitors but that evolve into something other than present-day massive red galaxies. Even though both stellar population synthesis models used in this study give an identical trend of this luminosity excess with magnitude, we find it to be 2-3 times larger with the M11 models. To help identify the reason for this luminosity excess, we look at the rate of change in luminosity per object, $r_g$, and at the evolution of the large-scale clustering.

The estimator $r_g$ was designed to be intrinsically less sensitive to contaminants in CMASS (provided they have a similar luminosity distribution to the galaxies of interest, in which case $r_g$ can be interpreted as a merger rate). It is also less sensitive the slope of the k+E corrections (mostly via the sample selection, which is different to the sample selection needed for $r_N$ and $r_\ell$ - see Section \ref{sec:sample_selection}). We find that $r_g$ tests differences in shape of the luminosity function, whilst $r_N$ and $r_\ell$ are also sensitive to the relative amplitudes. We only small evolution in $r_g$. Moreover, we find that this evolution (between $-1\%$ at the bright end and $2\%$ at the faint end - see left-hand panel of Fig.~\ref{fig:r_g}) is much less sensitive to the stellar population synthesis modelling. We compared the optimally \vmatch-weighted $r_g$ with the results obtained using a standard $V/$\vmax weight (Fig.~\ref{fig:r_g}). In that case we would infer a merger rate of up to $13\%$ at the faintest end, showing that the weighting scheme we introduce is having the effect we intended: effectively matching the two samples terms of comoving densities and average luminosity per galaxy. This result is evidence for a small evolution in the properties of the galaxies, once they are properly weighted. 

We should comment on a possible systematic error in $r_g$ due to the completeness estimation in each survey. In Section \ref{sec:luminosity_function} we noted that the fraction of CMASS galaxies that fail the star-galaxy separation (but which are nonetheless genuine galaxies) is larger than the fraction of LRGs that fail similar cuts at low redshift by at most $1\%$. This is a small number, but comparable to the accuracy with which we aim to constrain the dynamical evolution of these galaxies. We should therefore conservatively add $1\%$ uncertainty to the rates we show in Section \ref{sec:population_growth}. For our most inclusive sample ($M_{i 0.55} < -22$), $r_g$ is therefore constrained to be $1.6 \pm 1.5\%$ with M11 models, and $0.4 \pm 1.4\%$ with FSPS models. Our most inclusive samples include $\approx 95\%$ of the LRGs and $\approx 40\%$ of CMASS galaxies, and have large stellar masses with $\log_{10} M/M_{\odot} \gtrsim 11.2$ (Maraston et al. 2012, in prep).

To place further constraints on the evolution of our weighted samples, we investigate the evolution of the amplitude of the large-scale clustering (Fig.~\ref{fig:xi0_ev}). We find that the best fit to a dynamically passive evolution model (i.e. strictly no mergers, and assuming LCDM) happens when we weight the galaxies by \vmatch and luminosity, and we use only red galaxies ($g-i > 2.35$). This is {\it further} evidence that the weights are working as they should, and we find formally good fits to the passive model using both FSPS and M11 solutions, providing increased weight to our interpretation of Figs.~\ref{fig:luminosity_growth}, \ref{fig:merger_rates} and \ref{fig:r_g}.


\subsection{Comparison with previous work}

Independent efforts to constrain the assembly and evolution of massive galaxies range generally focus on the cosmic evolution of the luminosity function or on their clustering. Direct comparisons are difficult due to the different samples used, which vary in terms of number density (or typical minimum halo mass), magnitude or colour cuts, but this is nonetheless a useful exercise.

\cite{CoolEtAl08} measured the evolution in the luminosity function of LRGs up to $z\approx 0.9$ by observing a small sample of nearly 300 galaxies targeted using similar cuts to SDSS-II LRGs. They find that at the massive end, galaxies must not have grown by more than 50\% between $z=0.9$ and $z=0.1$, corresponding to less than 8\% growth per Gyr. \cite{WakeEtAl06} analysed SDSS-I/II and 2SLAQ LRGs, covering a redshift range between $z=0.17$ and $z=0.6$ and found that no growth was necessary to explain the evolution of the luminosity function once a passive stellar evolution model was applied, but that non-passive growth with up to 25\% of the galaxies merging between the two redshifts could not be ruled out. \cite{BrownEtAl07} studied the luminosity evolution of nearly 40,000 galaxies since $z\approx1$ in the Bootes field. At the massive red end, they find that at least 80\% of the stellar mass of these galaxies had to be place by $z=0.9$, resulting in a mild growth of less than $3\%$ per Gyr to the present-day. \cite{TojeiroEtAl10} measured the stellar growth of LRGs between $0.2\lesssim z \lesssim 0.45$ and found it to vary between 2-6\%, depending on the luminosity with brighter galaxies showing a smaller evolution. For the same luminosity range in this paper we find a smaller merger rate, in spite of the larger redshift range. This discrepancy is explained by a combination of different model for the stellar evolution of the galaxies and by an estimator that is less sensitive to overall offsets in magnitudes and potential contaminants (see equation \ref{eq:r_g}). Whereas studying the evolution of luminosity densities can certainly test departure from a purely or nearly passive dynamical evolution, it only formally tests a net influx of luminosity in or out of the samples between the two redshifts. Studies of the evolution of the clustering of galaxies are therefore important in characterising the mass assembly of these massive galaxies.

By fitting a halo model to a sample at high-redshift, evolving it passively to a lower redshift, and comparing it directly to a model fitted to data at that redshift allows for a direct comparison of predicted (from a passive model) and measured populations of centrals and satellites for samples matched in number density. Traditionally such approaches find too many satellites at low redshift, resulting in too large a number density and an overestimation of power at all scales (see e.g. \citealt{WhiteEtAl07, BrownEtAl08, WakeEtAl08}). 

The usual interpretation is that a fraction of the satellites and/or centrals must have merged. Such approaches have implied merger rates in the order of 2-8\% per Gyr since a redshift of 1, which is in apparent agreement with the results presented here. An interesting disagreement, however, is the fact that such studies have found that the evolution of the large-scale power departs significantly from the passive model of \cite{Fry96}; they find a nearly invariant large-scale amplitude as a function of redshift, which corresponds to effectively {\em underestimating} the bias at low redshift, with respect to passive evolution. As mentioned previously the merging scenario - coming from HOD fits to smaller scales - suggests that a fraction of galaxies in the sample at high-redshift must have merged, effectively reducing the large-scale power via two mechanisms. It primarily reduces the number of objects in high-mass halos, therefore decreasing the overall bias of the sample. A likely secondary effect is that merging within the sample must reduce the number density at low redshift for the same population of galaxies - as the samples are matched in number density, resulting in an enrichment of less-biased galaxies at low redshift. 

Our matching and weighting by luminosity bypases both of these problems. Firstly, when weighting galaxies by their luminosity, merging events between galaxies will not decrease the relative contribution of the halos within which they reside to the overall bias of the sample. This is only strictly true in the case of no loss of light to the ICM, but we argue here that weighting by luminosity will almost always be better than any weighting scheme that depends on the number of objects - in a merging of two objects the relative contribution of a given halo will be reduced by $1/2$ if weighting by number. It follows that, provided that the overall loss of light is less than $50\%$ of the combined light of the merging system, we have an estimation of the bias evolution that is less sensitive to merging of galaxies within the sample, and to which the Fry model is more applicable. Note that this is true {\em even} in the case of merging within the sample. Estimates on how much light may be lost to the intra-cluster medium varies. Based on a halo model analysis, \cite{WhiteEtAl07} estimated an upper bound on the loss of light to the ICM of $25-40\%$ of the light of the accreting satellite (not of the whole merging system), depending on halo mass; \cite{SkibbaEtAl07} argued that this fraction should be between $5-15\%$.  Using simulation-based models \cite{ConroyEtAl07b} show how scenarios that allow disrupted satellites to deposit up to 80\% of their stars in the ICM favour the observed evolution in the galaxy stellar mass function since $z\sim1$. \cite{PurcellEtAl07} estimate that around 20\% of the total light in massive halos ($M>10^{13}M_\odot$) is in a diffuse component. Analyses as the ones above show how understanding the mechanisms that lead to the formation of the ICL is of clear importance to learning how massive galaxies assemble. Nonetheless it remains clear that weighting by luminosity is {\em in practice} advantageous if one wishes, as we do, to use a passive model for the evolution of the large-scale bias. The effectiveness of our weighting scheme is nicely demonstrated  with our clustering analysis, which shows how a passive model is a better fit to the weighted data (see Fig.~\ref{fig:xi0_ev}). 

\subsection{Final remarks and future work}
We conclude that our sample is slowly evolving (to less that $2\%$ by merging, when samples are appropriately matched and weighted), to the extent of what is testable by current data and models. We demonstrated the efficiency of a \vmatch and luminosity weighting in constructing a sample that is as close to being dynamically passive as possible, whilst at the same time not needing to cut the sample in colour or redshift; therefore optimising our signal. This aspect is particularly important given broad nature of CMASS galaxies with respect to LRGs.

In terms of future work, we will assess the cosmological gains of using our slowly evolving sample when measuring growth rates and redshifts-space distortions using the large-scale amplitude of the correlation function. We further intend to extend this analysis to the LOZ and main galaxy samples \citep{StraussEtAl02}, which will provide not only better statistics for cosmology analysis, but will also allow a study of evolutionary paths of other populations of galaxies.


\section{Acknowledgments}
RT and WJP thank the European Research Council for support. WJP also thanks the Science and Technology Facilities Council. 

Funding for SDSS-III has been provided by the Alfred P. Sloan Foundation, the Participating Institutions, the National Science Foundation, and the U.S. Department of Energy. The SDSS-III web site is http://www.sdss3.org/.

SDSS-III is managed by the Astrophysical Research Consortium for the
Participating Institutions of the SDSS-III Collaboration including the
University of Arizona,
the Brazilian Participation Group,
Brookhaven National Laboratory,
University of Cambridge,
Carnegie Mellon University,
University of Florida,
the French Participation Group,
the German Participation Group,
Harvard University,
the Instituto de Astrofisica de Canarias,
the Michigan State/Notre Dame/JINA Participation Group,
Johns Hopkins University,
Lawrence Berkeley National Laboratory,
Max Planck Institute for Astrophysics,
Max Planck Institute for Extraterrestrial Physics,
New Mexico State University,
New York University,
Ohio State University,
Pennsylvania State University,
University of Portsmouth,
Princeton University,
the Spanish Participation Group,
University of Tokyo,
University of Utah,
Vanderbilt University,
University of Virginia,
University of Washington,
and Yale University.
\bibliographystyle{mn2e}
\bibliography{my_bibliography}

\begin{thebibliography}{}

\bibitem[\protect\citeauthoryear{{Abazajian} et~al.,}{{Abazajian}
  et~al.}{2009}]{AbazajianEtAl09}
{Abazajian} K.~N.,  et~al., 2009, \apjs, 182, 543

\bibitem[\protect\citeauthoryear{{Aihara} et~al.,}{{Aihara}
  et~al.}{2011}]{AiharaEtAl11}
{Aihara} H.,  et~al., 2011, \apjs, 193, 29

\bibitem[\protect\citeauthoryear{{Anderson} et~al.,}{{Anderson}
  et~al.}{2012}]{Aadvark}
{Anderson} L.,  et~al., 2012, ArXiv e-prints

\bibitem[\protect\citeauthoryear{{Bell}, {Phleps}, {Somerville}, {Wolf},
  {Borch} \& {Meisenheimer}}{{Bell} et~al.}{2006}]{BellEtAl06}
{Bell} E.~F.,  {Phleps} S.,  {Somerville} R.~S.,  {Wolf} C.,  {Borch} A.,
  {Meisenheimer} K.,  2006, \apj, 652, 270

\bibitem[\protect\citeauthoryear{{Bernardi}, {Roche}, {Shankar} \&
  {Sheth}}{{Bernardi} et~al.}{2011}]{BernardiEtAl10b}
{Bernardi} M.,  {Roche} N.,  {Shankar} F.,    {Sheth} R.~K.,  2011, \mnras,
  412, L6

\bibitem[\protect\citeauthoryear{{Brown}, {Dey}, {Jannuzi}, {Brand}, {Benson},
  {Brodwin}, {Croton} \& {Eisenhardt}}{{Brown} et~al.}{2007}]{BrownEtAl07}
{Brown} M.~J.~I.,  {Dey} A.,  {Jannuzi} B.~T.,  {Brand} K.,  {Benson} A.~J.,
  {Brodwin} M.,  {Croton} D.~J.,    {Eisenhardt} P.~R.,  2007, \apj, 654, 858

\bibitem[\protect\citeauthoryear{{Brown}, {Zheng}, {White}, {Dey}, {Jannuzi},
  {Benson}, {Brand}, {Brodwin} \& {Croton}}{{Brown} et~al.}{2008}]{BrownEtAl08}
{Brown} M.~J.~I.,  {Zheng} Z.,  {White} M.,  {Dey} A.,  {Jannuzi} B.~T.,
  {Benson} A.~J.,  {Brand} K.,  {Brodwin} M.,    {Croton} D.~J.,  2008, \apj,
  682, 937

\bibitem[\protect\citeauthoryear{{Bundy}, {Fukugita}, {Ellis}, {Targett},
  {Belli} \& {Kodama}}{{Bundy} et~al.}{2009}]{BundyEtAl09}
{Bundy} K.,  {Fukugita} M.,  {Ellis} R.~S.,  {Targett} T.~A.,  {Belli} S.,
  {Kodama} T.,  2009, \apj, 697, 1369

\bibitem[\protect\citeauthoryear{{Cannon} et~al.,}{{Cannon}
  et~al.}{2006}]{CannonEtAl06}
{Cannon} R.,  et~al., 2006, \mnras, 372, 425

\bibitem[\protect\citeauthoryear{{Carson} \& {Nichol}}{{Carson} \&
  {Nichol}}{2010}]{CarsonEtAl10}
{Carson} D.~P.,  {Nichol} R.~C.,  2010, \mnras, pp 1163--+

\bibitem[\protect\citeauthoryear{{Cassisi}, {Castellani} \&
  {Castellani}}{{Cassisi} et~al.}{1997}]{CassisiEtAl97}
{Cassisi} S.,  {Castellani} M.,    {Castellani} V.,  1997, \aap, 317, 108

\bibitem[\protect\citeauthoryear{{Chabrier}}{{Chabrier}}{2003}]{Chabrier03}
{Chabrier} G.,  2003, \pasp, 115, 763

\bibitem[\protect\citeauthoryear{{Charlot} \& {Fall}}{{Charlot} \&
  {Fall}}{2000}]{CharlotFall00}
{Charlot} S.,  {Fall} S.~M.,  2000, \apj, 539, 718

\bibitem[\protect\citeauthoryear{{Conroy} \& {Gunn}}{{Conroy} \&
  {Gunn}}{2010}]{ConroyAndGunn10}
{Conroy} C.,  {Gunn} J.~E.,  2010, \apj, 712, 833

\bibitem[\protect\citeauthoryear{{Conroy}, {Gunn} \& {White}}{{Conroy}
  et~al.}{2009}]{ConroyEtAl09}
{Conroy} C.,  {Gunn} J.~E.,    {White} M.,  2009, \apj, 699, 486

\bibitem[\protect\citeauthoryear{{Conroy}, {Ho} \& {White}}{{Conroy}
  et~al.}{2007}]{ConroyEtAl07a}
{Conroy} C.,  {Ho} S.,    {White} M.,  2007, \mnras, 379, 1491

\bibitem[\protect\citeauthoryear{{Conroy}, {Wechsler} \& {Kravtsov}}{{Conroy}
  et~al.}{2007}]{ConroyEtAl07b}
{Conroy} C.,  {Wechsler} R.~H.,    {Kravtsov} A.~V.,  2007, \apj, 668, 826

\bibitem[\protect\citeauthoryear{{Cool}, {Eisenstein}, {Johnston}, {Scranton},
  {Brinkmann}, {Schneider} \& {Zehavi}}{{Cool} et~al.}{2006}]{CoolEtAl06}
{Cool} R.~J.,  {Eisenstein} D.~J.,  {Johnston} D.,  {Scranton} R.,  {Brinkmann}
  J.,  {Schneider} D.~P.,    {Zehavi} I.,  2006, \aj, 131, 736

\bibitem[\protect\citeauthoryear{{Cool} et~al.,}{{Cool}
  et~al.}{2008}]{CoolEtAl08}
{Cool} R.~J.,  et~al., 2008, \apj, 682, 919

\bibitem[\protect\citeauthoryear{{Cresswell} \& {Percival}}{{Cresswell} \&
  {Percival}}{2009}]{CresswellEtAl09}
{Cresswell} J.~G.,  {Percival} W.~J.,  2009, \mnras, 392, 682

\bibitem[\protect\citeauthoryear{{De Propris} et~al.,}{{De Propris}
  et~al.}{2010}]{deProprisEtAl10}
{De Propris} R.,  et~al., 2010, \aj, 139, 794

\bibitem[\protect\citeauthoryear{{Eisenstein} et~al.,}{{Eisenstein}
  et~al.}{2001}]{EisensteinEtAl01}
{Eisenstein} D.~J.,  et~al., 2001, \aj, 122, 2267

\bibitem[\protect\citeauthoryear{{Eisenstein} et~al.,}{{Eisenstein}
  et~al.}{2003}]{EisensteinEtAl03}
{Eisenstein} D.~J.,  et~al., 2003, \apj, 585, 694

\bibitem[\protect\citeauthoryear{{Eisenstein} et~al.,}{{Eisenstein}
  et~al.}{2011}]{EisensteinEtAl11}
{Eisenstein} D.~J.,  et~al., 2011, \aj, 142, 72

\bibitem[\protect\citeauthoryear{{Faber} et~al.,}{{Faber}
  et~al.}{2007}]{FaberEtAl07}
{Faber} S.~M.,  et~al., 2007, \apj, 665, 265

\bibitem[\protect\citeauthoryear{{Feldmeier}, {Mihos}, {Morrison}, {Harding},
  {Kaib} \& {Dubinski}}{{Feldmeier} et~al.}{2004}]{FeldmeierEtAl04}
{Feldmeier} J.~J.,  {Mihos} J.~C.,  {Morrison} H.~L.,  {Harding} P.,  {Kaib}
  N.,    {Dubinski} J.,  2004, \apj, 609, 617

\bibitem[\protect\citeauthoryear{{Fry}}{{Fry}}{1996}]{Fry96}
{Fry} J.~N.,  1996, \apjl, 461, L65+

\bibitem[\protect\citeauthoryear{{Fukugita}, {Ichikawa}, {Gunn}, {Doi},
  {Shimasaku} \& {Schneider}}{{Fukugita} et~al.}{1996}]{FukugitaEtAl96}
{Fukugita} M.,  {Ichikawa} T.,  {Gunn} J.~E.,  {Doi} M.,  {Shimasaku} K.,
  {Schneider} D.~P.,  1996, \aj, 111, 1748

\bibitem[\protect\citeauthoryear{{Graham}, {Driver}, {Petrosian}, {Conselice},
  {Bershady}, {Crawford} \& {Goto}}{{Graham} et~al.}{2005}]{GrahamEtAl05}
{Graham} A.~W.,  {Driver} S.~P.,  {Petrosian} V.,  {Conselice} C.~J.,
  {Bershady} M.~A.,  {Crawford} S.~M.,    {Goto} T.,  2005, \aj, 130, 1535

\bibitem[\protect\citeauthoryear{{Graves}, {Faber} \& {Schiavon}}{{Graves}
  et~al.}{2009}]{GravesEtAl09}
{Graves} G.~J.,  {Faber} S.~M.,    {Schiavon} R.~P.,  2009, \apj, 698, 1590

\bibitem[\protect\citeauthoryear{{Gunn} et~al.,}{{Gunn}
  et~al.}{1998}]{GunnEtAl98}
{Gunn} J.~E.,  et~al., 1998, \aj, 116, 3040

\bibitem[\protect\citeauthoryear{{Gunn} et~al.,}{{Gunn}
  et~al.}{2006}]{GunnEtAl06}
{Gunn} J.~E.,  et~al., 2006, \aj, 131, 2332

\bibitem[\protect\citeauthoryear{{Hamilton}}{{Hamilton}}{1992}]{Hamilton92}
{Hamilton} A.~J.~S.,  1992, \apjl, 385, L5

\bibitem[\protect\citeauthoryear{{Jimenez}, {Bernardi}, {Haiman}, {Panter} \&
  {Heavens}}{{Jimenez} et~al.}{2007}]{JimenezEtAl07}
{Jimenez} R.,  {Bernardi} M.,  {Haiman} Z.,  {Panter} B.,    {Heavens} A.~F.,
  2007, \apj, 669, 947

\bibitem[\protect\citeauthoryear{{Kaviraj} et~al.,}{{Kaviraj}
  et~al.}{2007}]{KavirajEtAl07}
{Kaviraj} S.,  et~al., 2007, \apjs, 173, 619

\bibitem[\protect\citeauthoryear{{Kaviraj}, {Peirani}, {Khochfar}, {Silk} \&
  {Kay}}{{Kaviraj} et~al.}{2009}]{KavirajEtAl09}
{Kaviraj} S.,  {Peirani} S.,  {Khochfar} S.,  {Silk} J.,    {Kay} S.,  2009,
  \mnras, 394, 1713

\bibitem[\protect\citeauthoryear{{Kaviraj}, {Tan}, {Ellis} \& {Silk}}{{Kaviraj}
  et~al.}{2010}]{KavirajEtAl10}
{Kaviraj} S.,  {Tan} K.,  {Ellis} R.~S.,    {Silk} J.,  2010, ArXiv e-prints

\bibitem[\protect\citeauthoryear{{Kroupa}}{{Kroupa}}{2001}]{Kroupa01}
{Kroupa} P.,  2001, \mnras, 322, 231

\bibitem[\protect\citeauthoryear{{Landy} \& {Szalay}}{{Landy} \&
  {Szalay}}{1993}]{LandySzalay93}
{Landy} S.~D.,  {Szalay} A.~S.,  1993, \apj, 412, 64

\bibitem[\protect\citeauthoryear{{Manera} et~al.,}{{Manera}
  et~al.}{2012}]{ManeraEtAl12}
{Manera} M.,  et~al., 2012, ArXiv e-prints: 1203.6609

\bibitem[\protect\citeauthoryear{{Maraston} \& {Str{\"o}mb{\"a}ck}}{{Maraston}
  \& {Str{\"o}mb{\"a}ck}}{2011}]{MarastonEtAl11}
{Maraston} C.,  {Str{\"o}mb{\"a}ck} G.,  2011, \mnras, 418, 2785

\bibitem[\protect\citeauthoryear{{Maraston}, {Str{\"o}mb{\"a}ck}, {Thomas},
  {Wake} \& {Nichol}}{{Maraston} et~al.}{2009}]{MarastonEtAl09}
{Maraston} C.,  {Str{\"o}mb{\"a}ck} G.,  {Thomas} D.,  {Wake} D.~A.,
  {Nichol} R.~C.,  2009, \mnras, 394, L107

\bibitem[\protect\citeauthoryear{{Marigo} \& {Girardi}}{{Marigo} \&
  {Girardi}}{2007}]{MarigoEtAl07}
{Marigo} P.,  {Girardi} L.,  2007, \aaps, 469, 239

\bibitem[\protect\citeauthoryear{{Marigo}, {Girardi}, {Bressan}, {Groenewegen},
  {Silva} \& {Granato}}{{Marigo} et~al.}{2008}]{MarigoEtAl08}
{Marigo} P.,  {Girardi} L.,  {Bressan} A.,  {Groenewegen} M.~A.~T.,  {Silva}
  L.,    {Granato} G.~L.,  2008, \aaps, 482, 883

\bibitem[\protect\citeauthoryear{{Masjedi} et~al.,}{{Masjedi}
  et~al.}{2006}]{MasjediEtAl06}
{Masjedi} M.,  et~al., 2006, \apj, 644, 54

\bibitem[\protect\citeauthoryear{{Masjedi}, {Hogg} \& {Blanton}}{{Masjedi}
  et~al.}{2008}]{MasjediEtAl08}
{Masjedi} M.,  {Hogg} D.~W.,    {Blanton} M.~R.,  2008, \apj, 679, 260

\bibitem[\protect\citeauthoryear{{Masters} et~al.,}{{Masters}
  et~al.}{2011}]{MastersEtAl11}
{Masters} K.~L.,  et~al., 2011, \mnras, pp 1417--+

\bibitem[\protect\citeauthoryear{{Mihos}, {Harding}, {Feldmeier} \&
  {Morrison}}{{Mihos} et~al.}{2005}]{MihosEtAl05}
{Mihos} J.~C.,  {Harding} P.,  {Feldmeier} J.,    {Morrison} H.,  2005, \apjl,
  631, L41

\bibitem[\protect\citeauthoryear{{Padmanabhan} et~al.,}{{Padmanabhan}
  et~al.}{2008}]{PadmanabhanEtAl98}
{Padmanabhan} N.,  et~al., 2008, \apj, 674, 1217

\bibitem[\protect\citeauthoryear{{Percival} et~al.,}{{Percival}
  et~al.}{2007}]{PercivalEtAl07}
{Percival} W.~J.,  et~al., 2007, \apj, 657, 51

\bibitem[\protect\citeauthoryear{{Petrosian}}{{Petrosian}}{1976}]{Petrosian76}
{Petrosian} V.,  1976, \apjl, 209, L1

\bibitem[\protect\citeauthoryear{{Polarski} \& {Gannouji}}{{Polarski} \&
  {Gannouji}}{2008}]{PolarskiEtAl08}
{Polarski} D.,  {Gannouji} R.,  2008, Physics Letters B, 660, 439

\bibitem[\protect\citeauthoryear{{Purcell}, {Bullock} \& {Zentner}}{{Purcell}
  et~al.}{2007}]{PurcellEtAl07}
{Purcell} C.~W.,  {Bullock} J.~S.,    {Zentner} A.~R.,  2007, \apj, 666, 20

\bibitem[\protect\citeauthoryear{{Reid} et~al.,}{{Reid}
  et~al.}{2012}]{ReidEtAl12}
{Reid} B.~A.,  et~al., 2012, ArXiv e-prints:1203.6641

\bibitem[\protect\citeauthoryear{{Renzini} \& {Buzzoni}}{{Renzini} \&
  {Buzzoni}}{1986}]{RenziniAndBuzzoni86}
{Renzini} A.,  {Buzzoni} A.,  1986, in {C.~Chiosi \& A.~Renzini} ed., Spectral
  Evolution of Galaxies Vol.~122 of Astrophysics and Space Science Library,
  {Global properties of stellar populations and the spectral evolution of
  galaxies}.
pp 195--231

\bibitem[\protect\citeauthoryear{{Ross} \& {Brunner}}{{Ross} \&
  {Brunner}}{2009}]{RossEtAl09}
{Ross} A.~J.,  {Brunner} R.~J.,  2009, \mnras, 399, 878

\bibitem[\protect\citeauthoryear{{Ross} et~al.,}{{Ross}
  et~al.}{2011}]{RossEtAl11b}
{Ross} A.~J.,  et~al., 2011, \mnras, 417, 1350

\bibitem[\protect\citeauthoryear{{Ross} et~al.,}{{Ross}
  et~al.}{2012}]{RossEtAl12}
{Ross} A.~J.,  et~al., 2012, ArXiv e-prints: 1203.6499

\bibitem[\protect\citeauthoryear{{Ross}, {Percival} \& {Brunner}}{{Ross}
  et~al.}{2010}]{RossEtAl10}
{Ross} A.~J.,  {Percival} W.~J.,    {Brunner} R.~J.,  2010, \mnras, 407, 420

\bibitem[\protect\citeauthoryear{{Ross}, {Tojeiro} \& {Percival}}{{Ross}
  et~al.}{2011}]{RossEtAl11}
{Ross} A.~J.,  {Tojeiro} R.,    {Percival} W.~J.,  2011, \mnras, 413, 2078

\bibitem[\protect\citeauthoryear{{Salim} \& {Rich}}{{Salim} \&
  {Rich}}{2010}]{SalimEtAl10}
{Salim} S.,  {Rich} R.~M.,  2010, \apjl, 714, L290

\bibitem[\protect\citeauthoryear{{Samushia}, {Percival} \&
  {Raccanelli}}{{Samushia} et~al.}{2012}]{SamushiaEtAl12}
{Samushia} L.,  {Percival} W.~J.,    {Raccanelli} A.,  2012, \mnras, 420, 2102

\bibitem[\protect\citeauthoryear{{S{\'a}nchez-Bl{\'a}zquez}, {Peletier},
  {Jim{\'e}nez-Vicente}, {Cardiel}, {Cenarro}, {Falc{\'o}n-Barroso}, {Gorgas},
  {Selam} \& {Vazdekis}}{{S{\'a}nchez-Bl{\'a}zquez}
  et~al.}{2006}]{SanchezBlazquezEtAl06}
{S{\'a}nchez-Bl{\'a}zquez} P.,  {Peletier} R.~F.,  {Jim{\'e}nez-Vicente} J.,
  {Cardiel} N.,  {Cenarro} A.~J.,  {Falc{\'o}n-Barroso} J.,  {Gorgas} J.,
  {Selam} S.,    {Vazdekis} A.,  2006, \mnras, 371, 703

\bibitem[\protect\citeauthoryear{{Schaller}, {Schaerer}, {Meynet} \&
  {Maeder}}{{Schaller} et~al.}{1992}]{SchallerEtAl92}
{Schaller} G.,  {Schaerer} D.,  {Meynet} G.,    {Maeder} A.,  1992, \aaps, 96,
  269

\bibitem[\protect\citeauthoryear{{Schawinski} et~al.,}{{Schawinski}
  et~al.}{2007}]{SchawinskiEtAl07}
{Schawinski} K.,  et~al., 2007, \apjs, 173, 512

\bibitem[\protect\citeauthoryear{{Schlegel}, {Finkbeiner} \&
  {Davis}}{{Schlegel} et~al.}{1998}]{SchlegelEtAl98}
{Schlegel} D.~J.,  {Finkbeiner} D.~P.,    {Davis} M.,  1998, \apj, 500, 525

\bibitem[\protect\citeauthoryear{{Seo}, {Eisenstein} \& {Zehavi}}{{Seo}
  et~al.}{2008}]{SeoEtAl08}
{Seo} H.-J.,  {Eisenstein} D.~J.,    {Zehavi} I.,  2008, \apj, 681, 998

\bibitem[\protect\citeauthoryear{{Sheth}, {Jimenez}, {Panter} \&
  {Heavens}}{{Sheth} et~al.}{2006}]{ShethEtAl06}
{Sheth} R.~K.,  {Jimenez} R.,  {Panter} B.,    {Heavens} A.~F.,  2006, \apjl,
  650, L25

\bibitem[\protect\citeauthoryear{{Skibba}}{{Skibba}}{2009}]{SkibbaEtAl09b}
{Skibba} R.~A.,  2009, \mnras, 392, 1467

\bibitem[\protect\citeauthoryear{{Skibba} \& {Sheth}}{{Skibba} \&
  {Sheth}}{2009}]{SkibbaEtAl09a}
{Skibba} R.~A.,  {Sheth} R.~K.,  2009, \mnras, 392, 1080

\bibitem[\protect\citeauthoryear{{Skibba}, {Sheth} \& {Martino}}{{Skibba}
  et~al.}{2007}]{SkibbaEtAl07}
{Skibba} R.~A.,  {Sheth} R.~K.,    {Martino} M.~C.,  2007, \mnras, 382, 1940

\bibitem[\protect\citeauthoryear{{Stoughton} et~al.,}{{Stoughton}
  et~al.}{2002}]{StoughtonEtAl02}
{Stoughton} C.,  et~al., 2002, \aj, 123, 485

\bibitem[\protect\citeauthoryear{{Strauss} et~al.,}{{Strauss}
  et~al.}{2002}]{StraussEtAl02}
{Strauss} M.~A.,  et~al., 2002, \aj, 124, 1810

\bibitem[\protect\citeauthoryear{{Swanson}, {Tegmark}, {Blanton} \&
  {Zehavi}}{{Swanson} et~al.}{2008}]{SwansonEtAl08}
{Swanson} M.~E.~C.,  {Tegmark} M.,  {Blanton} M.,    {Zehavi} I.,  2008,
  \mnras, 385, 1635

\bibitem[\protect\citeauthoryear{{Tal}, {Wake}, {van Dokkum}, {van den Bosch},
  {Schneider}, {Brinkmann} \& {Weaver}}{{Tal} et~al.}{2012}]{TalEtAl12}
{Tal} T.,  {Wake} D.~A.,  {van Dokkum} P.~G.,  {van den Bosch} F.~C.,
  {Schneider} D.~P.,  {Brinkmann} J.,    {Weaver} B.~A.,  2012, \apj, 746, 138

\bibitem[\protect\citeauthoryear{{Tegmark} \& {Peebles}}{{Tegmark} \&
  {Peebles}}{1998}]{TegmarkEtAl98}
{Tegmark} M.,  {Peebles} P.~J.~E.,  1998, \apjl, 500, L79

\bibitem[\protect\citeauthoryear{{Thomas}, {Maraston}, {Bender} \& {Mendes de
  Oliveira}}{{Thomas} et~al.}{2005}]{ThomasEtAl05}
{Thomas} D.,  {Maraston} C.,  {Bender} R.,    {Mendes de Oliveira} C.,  2005,
  \apj, 621, 673

\bibitem[\protect\citeauthoryear{{Thomas}, {Maraston}, {Schawinski}, {Sarzi} \&
  {Silk}}{{Thomas} et~al.}{2010}]{ThomasEtAl10}
{Thomas} D.,  {Maraston} C.,  {Schawinski} K.,  {Sarzi} M.,    {Silk} J.,
  2010, \mnras, 404, 1775

\bibitem[\protect\citeauthoryear{{Tinker} \& {Wetzel}}{{Tinker} \&
  {Wetzel}}{2010}]{TinkerEtAl10}
{Tinker} J.~L.,  {Wetzel} A.~R.,  2010, \apj, 719, 88

\bibitem[\protect\citeauthoryear{{Tojeiro}, {Heavens}, {Jimenez} \&
  {Panter}}{{Tojeiro} et~al.}{2007}]{TojeiroEtAl07}
{Tojeiro} R.,  {Heavens} A.~F.,  {Jimenez} R.,    {Panter} B.,  2007, \mnras,
  381, 1252

\bibitem[\protect\citeauthoryear{{Tojeiro} \& {Percival}}{{Tojeiro} \&
  {Percival}}{2010}]{TojeiroEtAl10}
{Tojeiro} R.,  {Percival} W.~J.,  2010, \mnras, 405, 2534

\bibitem[\protect\citeauthoryear{{Tojeiro} \& {Percival}}{{Tojeiro} \&
  {Percival}}{2011}]{TojeiroEtAl11b}
{Tojeiro} R.,  {Percival} W.~J.,  2011, \mnras, 417, 1114

\bibitem[\protect\citeauthoryear{{Tojeiro}, {Percival}, {Heavens} \&
  {Jimenez}}{{Tojeiro} et~al.}{2011}]{TojeiroEtAl11}
{Tojeiro} R.,  {Percival} W.~J.,  {Heavens} A.~F.,    {Jimenez} R.,  2011,
  \mnras, 413, 434

\bibitem[\protect\citeauthoryear{{Tojeiro}, {Wilkins}, {Heavens}, {Panter} \&
  {Jimenez}}{{Tojeiro} et~al.}{2009}]{TojeiroEtAl09}
{Tojeiro} R.,  {Wilkins} S.,  {Heavens} A.~F.,  {Panter} B.,    {Jimenez} R.,
  2009, \apjs, 185, 1

\bibitem[\protect\citeauthoryear{{Trager}, {Faber}, {Worthey} \&
  {Gonz{\'a}lez}}{{Trager} et~al.}{2000}]{TragerEtAl00}
{Trager} S.~C.,  {Faber} S.~M.,  {Worthey} G.,    {Gonz{\'a}lez} J.~J.,  2000,
  \aj, 120, 165

\bibitem[\protect\citeauthoryear{{Wake} et~al.,}{{Wake}
  et~al.}{2006}]{WakeEtAl06}
{Wake} D.~A.,  et~al., 2006, \mnras, 372, 537

\bibitem[\protect\citeauthoryear{{Wake} et~al.,}{{Wake}
  et~al.}{2008}]{WakeEtAl08}
{Wake} D.~A.,  et~al., 2008, \mnras, 387, 1045

\bibitem[\protect\citeauthoryear{{Wake} et~al.,}{{Wake}
  et~al.}{2011}]{WakeEtAl11}
{Wake} D.~A.,  et~al., 2011, \apj, 728, 46

\bibitem[\protect\citeauthoryear{{White} et~al.,}{{White}
  et~al.}{2011}]{WhiteEtAl11}
{White} M.,  et~al., 2011, \apj, 728, 126

\bibitem[\protect\citeauthoryear{{White}, {Zheng}, {Brown}, {Dey} \&
  {Jannuzi}}{{White} et~al.}{2007}]{WhiteEtAl07}
{White} M.,  {Zheng} Z.,  {Brown} M.~J.~I.,  {Dey} A.,    {Jannuzi} B.~T.,
  2007, \apjl, 655, L69

\bibitem[\protect\citeauthoryear{{Yang}, {Mo} \& {van den Bosch}}{{Yang}
  et~al.}{2009}]{YangEtAl09}
{Yang} X.,  {Mo} H.~J.,    {van den Bosch} F.~C.,  2009, \apj, 693, 830

\bibitem[\protect\citeauthoryear{{York} et~al.,}{{York}
  et~al.}{2000}]{YorkEtAl00}
{York} D.~G.,  et~al., 2000, \aj, 120, 1579

\bibitem[\protect\citeauthoryear{{Zehavi} et~al.,}{{Zehavi}
  et~al.}{2005a}]{ZehaviEtAl05a}
{Zehavi} I.,  et~al., 2005a, \apj, 621, 22

\bibitem[\protect\citeauthoryear{{Zehavi} et~al.,}{{Zehavi}
  et~al.}{2005b}]{ZehaviEtAl05b}
{Zehavi} I.,  et~al., 2005b, \apj, 630, 1

\bibitem[\protect\citeauthoryear{{Zehavi} et~al.,}{{Zehavi}
  et~al.}{2011}]{ZehaviEtAl11}
{Zehavi} I.,  et~al., 2011, \apj, 736, 59

\bibitem[\protect\citeauthoryear{{Zehavi}, {Patiri} \& {Zheng}}{{Zehavi}
  et~al.}{2012}]{ZehaviEtAl12}
{Zehavi} I.,  {Patiri} S.,    {Zheng} Z.,  2012, \apj, 746, 145

\bibitem[\protect\citeauthoryear{{Zheng}, {Coil} \& {Zehavi}}{{Zheng}
  et~al.}{2007}]{ZhengZEtAl07}
{Zheng} Z.,  {Coil} A.~L.,    {Zehavi} I.,  2007, \apj, 667, 760

\bibitem[\protect\citeauthoryear{{Zheng}, {Zehavi}, {Eisenstein}, {Weinberg} \&
  {Jing}}{{Zheng} et~al.}{2009}]{ZhengZEtAl09}
{Zheng} Z.,  {Zehavi} I.,  {Eisenstein} D.~J.,  {Weinberg} D.~H.,    {Jing}
  Y.~P.,  2009, \apj, 707, 554

\bibitem[\protect\citeauthoryear{{Zhu}, {Blanton} \& {Moustakas}}{{Zhu}
  et~al.}{2010}]{ZhuEtAl10}
{Zhu} G.,  {Blanton} M.~R.,    {Moustakas} J.,  2010, ArXiv e-prints

\end{thebibliography}

\end{document}